\documentclass[aps, prd, amsmath, floats, floatfix, twocolumn,superscriptaddress, nofootinbib, showpacs]{revtex4}

\usepackage{graphicx}
\usepackage{amsmath,amssymb}
\usepackage{amsfonts}
\usepackage{xspace} 
\usepackage[usenames]{color}
\usepackage{dcolumn}
\usepackage{bm}

\usepackage{ulem}
\newcommand{\Caltech}{\affiliation{Theoretical Astrophysics 350-17,
    California Institute of Technology, Pasadena, CA 91125}}
\newcommand{\Cornell}{\affiliation{Center for Radiophysics and Space
    Research, Cornell University, Ithaca, New York, 14853}}

\definecolor{darkgreen}{rgb}{0.2,0.7,0.2}

\def\gg{\mathfrak g} 

\newcommand{\TrExCurv}{K}    
\newcommand{\dtime}{\partial_t}   
\newcommand{\Lapse}{\alpha}       
\newcommand{\Shift}{\beta}   

\newcommand{\CTwoMetric}{\tilde{h}}         
\newcommand{\CSpatialNormal}{\tilde{s}}

\newcommand{\CF}{\psi}              

\newcommand{\CLapse}{\tilde{\alpha}}     
\newcommand{\dtCMetric}{\tilde{u
}}  

\newcommand{\CA}{\tilde{A}}         
\newcommand{\CRicciS}{\tilde{R}}    

\newcommand{\CCD}{{\tilde\nabla}\!} 
\newcommand{\CCDu}{{\tilde\nabla}}  
\newcommand{\CLong}[1]{(\tilde{\mathbb L}{#1})} 

\begin{document}

\title{Momentum flow in black-hole binaries: 
II. Numerical simulations of 
equal-mass, head-on mergers with antiparallel spins.}

\author{Geoffrey Lovelace} \Cornell %
\author{Yanbei Chen}   \Caltech %
\author{Michael Cohen} \Caltech %
\author{Jeffrey D. Kaplan}   \Caltech %
\author{Drew Keppel}   \Caltech %
\author{Keith D. Matthews} \Caltech %
\author{David A. Nichols}  \Caltech %
\author{Mark A. Scheel} \Caltech %
\author{Ulrich Sperhake} \Caltech %
\begin{abstract} 
Research on extracting science from binary-black-hole (BBH) simulations 
has often adopted 
a ``scattering matrix'' perspective: given the binary's initial 
parameters, what are the final hole's parameters and the emitted 
gravitational waveform? In contrast, we are using BBH
simulations to explore the nonlinear dynamics of curved spacetime.
Focusing on the head-on plunge, merger, and 
ringdown of a BBH with 
transverse, antiparallel spins, we explore 
numerically the momentum flow between the holes and the surrounding 
spacetime. We use the Landau-Lifshitz 
field-theory-in-flat-spacetime formulation
of general relativity to define and compute the density 
of 
field energy and field momentum outside horizons
and the energy and 
momentum contained within horizons, and 
we define the effective velocity of each
apparent and event
horizon as the ratio of its enclosed momentum to its enclosed mass-energy.
We find surprisingly good agreement between
the horizons' effective and coordinate velocities. 
During the plunge, 
the holes experience a frame-dragging-induced acceleration orthogonal
to the plane of their spins and their infall 
(``downward''), 
and they reach downward
speeds of order 1000 km/s. When the common apparent horizon 
forms (and when the event horizons merge and their merged neck expands), 
the horizon swallows upward field momentum that resided
between the holes, causing the merged hole 
to accelerate in 
the opposite (``upward'') direction. 
As the merged hole and the field energy and momentum settle 
down, a
pulsational burst of gravitational waves
is emitted, and the merged hole has a final 
effective velocity of about 20 km/s upward, 
which agrees with
the recoil velocity 
obtained by measuring
the linear momentum carried to infinity by the emitted gravitational 
radiation. 
To investigate the gauge dependence of our results, we compare 
pseudospectral and moving-puncture evolutions of 
physically similar initial data; 
although spectral and puncture simulations use different 
gauge conditions,
we find remarkably good 
agreement for our results in these two cases.
We also 
compare our simulations with the post-Newtonian trajectories and near-field 
energy-momentum. 
\end{abstract}

\date{\today \hspace{0.2truecm}}

\pacs{04.25.D-, 04.25.dg, 04.25.Nx, 
04.70.-s, 97.60.Lf}

\maketitle









\section{Introduction}
\label{sec:intro}
\subsection{Motivation}
\label{sec:intro_motivation}
Following Pretorius's 2005 breakthrough~\cite{Pretorius2005a},
several research groups have developed codes to 
solve Einstein's equations
numerically for
the inspiral, merger, and ringdown of colliding 
binary black holes (BBHs).  
Most simulations of BBH mergers to date 
have adopted
the moving-puncture method~\cite{Campanelli2006a,Baker2006a},
and 
spectral methods~\cite{Scheel2008} have also successfully simulated 
BBH mergers.

A major goal of current research is to successfully extract the physical
content of these simulations. Typically, efforts toward this goal 
adopt
a ``scattering matrix'' approach. 
Information obtained from numerical simulations on a finite set 
of
islands in the seven-dimensional\footnote{One parameter for the
mass ratio and six for the individual spins; additional parameters
might arise from eccentric orbits and the apparent dependence,
in at least some configurations,
of the recoil on the initial phase of the binary.}
parameter space is being extrapolated, by various research groups, 
to design  
complicated functions that give the final parameters of the merged hole 
and the
emitted gravitational waveforms as 
functions of the binary's initial parameters.


In this paper, 
however, we take a different perspective: we focus our attention
on the \emph{nonlinear dynamics of curved spacetime} 
during the holes' merger 
and ringdown.
Following Ref.~\cite{Chen2009} (paper I in this series),
our goal is to develop physical insight into the behavior of 
highly dynamical spacetimes such as the 
strong-field region
near the black-hole horizons in a merging binary.
As in paper I, we focus this study on the 
distribution and
flow of linear momentum in BBH spacetimes.
In contrast to
paper I's description of
the pre-merger motion of the holes in
the post-Newtonian approximation, in this paper we study the momentum flow 
during the plunge, merger, and ringdown of merging black holes in fully 
relativistic simulations.

\subsection{Linear momentum flow in BBHs and 
gauge dependence}
\label{sec:intro_momentumflow}
Typically, numerical simulations calculate only the 
\emph{total} linear momentum of a BBH system and
ignore the (gauge-dependent) linear momenta of the individual black holes.
However, linear momentum has been considered by 
Krishnan, Lousto and Zlochower~\cite{Krishnan:2007pu}.  
Inspired by the success of quasilocal 
angular momentum (see, e.g.,~\cite{Szabados2004} for a review) 
as a tool for measuring the spin of an 
individual black hole, 
Krishnan and colleagues proposed an analogous 
(but gauge-dependent) 
formula for the quasilocal linear momentum, and they calculate this 
quasilocal linear momentum for, e.g., the highly-spinning, unequal-mass 
BBH simulations 
in Ref.~\cite{LoustoZlochower2009}. 
This quasilocal linear momentum 
is also used to define an orbital angular momentum in 
Ref.~\cite{LoustoZlochower2008b}.

In this paper, we adopt a different, complementary method for measuring the 
holes' linear momenta: for the first time, we apply the Landau-Lifshitz 
momentum-flow formalism 
(described in paper I
and summarized in Sec.~\ref{sec:LLformal}) to 
numerical simulations of merging black holes. 
In this formalism, a mapping between the curved spacetime and 
an auxiliary flat spacetime (AFS) is chosen, and general relativity is 
reinterpreted as a field theory defined on this flat spacetime.
The AFS has a set of translational Killing vectors which we use 
to define a localized, conserved linear momentum. 
In particular, we
calculate i) a momentum density, ii) the momentum enclosed 
by horizons, and iii) the momentum enclosed by distant coordinate spheres. 
In the asymptotically flat region around a source, there is 
a preferred way to choose the mapping between the curved spacetime and the 
AFS; consequently, in this limit item iii) is gauge-invariant. 
In general, though, 
the choice of mapping is arbitrary, and it follows that
items i) and ii) are necessarily gauge-dependent. 

By examining the linear momentum flow in a dynamical 
spacetime---and living with the inevitable gauge dependence---we 
hope to develop strong intuition for the behavior of 
BBHs. As in paper I, we envision different 
numerical relativity groups choosing ``preferred'' gauges based on the 
coordinates of their numerical simulations.
While there is no reason, {\it a priori}, why simulations in different
gauges should agree, one of our hopes from paper I is realized 
\emph{for the
cases we consider}; namely,
in this paper, we 
calculate the horizon-enclosed momentum 
using spectral and moving-puncture 
evolutions of similar initial data,
and we do find surprisingly good agreement 
(cf. Figs.~\ref{fig:SpECVel} and~\ref{fig: punc_ll1}), 
even though the simulations use manifestly different gauge 
conditions [Eqs.~(\ref{eq:SpECGauge}) for the spectral simulations and 
Eqs. (\ref{eq:PunctureLapseGauge})--(\ref{eq: shift}) for the puncture 
simulations]. These are two of the most commonly used 
gauge conditions in numerical relativity.

Therefore, we continue to hope that in general---for the gauges 
commonly used in numerical simulations---the
momentum distributions for evolutions of physically similar initial 
data
will turn out to be at least qualitatively similar. 
If further 
investigation reveals this to be the case, then different 
research groups can simply use the coordinates 
used in the their simulations as the ``preferred coordinates'' for 
constructing the mapping to the AFS. 
Otherwise, we would advocate (as in Sec. I C of paper I)
that different numerical-relativity groups 
construct the mapping to the AFS by first agreeing
on a choice of ``preferred'' coordinates (e.g., a particular harmonic gauge)
and then 
transforming the results of their simulations to those coordinates.

\subsection{BBH mergers with recoil}
\label{sec:intro_recoil}
A particularly important application of this approach is 
an exploration of the momentum flow in BBH mergers with 
recoil.
The gravitational recoil or kick effect arising in a BBH
coalescence has attracted a great deal
of attention in recent years in the context of a variety of 
astrophysical scenarios including the structure of galaxies
\cite{BoylanKolchin2004, Gualandris2007, Komossa2008b},
the reionization history of the universe
\cite{Madau2004a},
the assembly of supermassive black holes \cite{Haiman2004,
Madau2004, Merritt2004, Volonteri2007b, Blecha2008} and
direct observational signatures
\cite{Loeb2007, Komossa2008, Menou2008}.
For a long time, estimates of the recoil magnitude were based
on approximative techniques \cite{Fitchett1983, Favata2004,
Blanchet2005b, Damour06}; accurate calculations in the
framework of fully nonlinear general relativity have only
become possible in the aftermath of important
breakthroughs in the field of numerical
relativity \cite{Pretorius2005a,Campanelli2006a,Baker2006a}.

Several groups have used numerical simulations
to study the kick resulting from the merger of
non-spinning and spinning binaries (see, e.g.,
\cite{Baker2006c,Gonzalez2007,Herrmann2007,Koppitz2007,Campanelli2007a,
Tichy:2007hk}). Most remarkably, recoil velocities of several
thousand km/s have been found for binaries with equal and opposite
spins in the orbital plane \cite{Campanelli2007a, Gonzalez2007b,
Campanelli2007}, and variants thereof with
hyperbolic orbits even generate $10^4~{\rm km/s}$
\cite{Healy2008}. Given the enormous astrophysical
repercussions of such large
recoil velocities, the community is now using various approaches
to obtain a better understanding of the kick as a function
of the initial BBH 
parameters~\cite{Boyle2007a,Boyle2007b,Schnittman2007a,Baker2008,
Tichy2008,Lousto2009}
resulting in phenomenological fitting 
formulas; see
\cite{Baker2007,LoustoZlochower2008b,Baker2008,Gonzalez2008,Rezzolla2008,
LoustoZlochower2009} and references therein.

On the other hand, 
our understanding of the local dynamics in these extraordinarily
violent events is still rather limited.
Some insight into the
origin of the holes' kick velocity has been obtained by examining the 
individual multipole moments of the emitted gravitational 
waves~\cite{Schnittman2007,MillerMatzner2008} and by 
approximating the recoil analytically using 
post-Newtonian~\cite{Blanchet2005b,Racine2008}, 
effective-one-body~\cite{Damour06}, and black-hole-perturbation 
theory~\cite{MinoBrink2008}. 
An intuitive
picture describing aspects of the so-called superkick
configurations generating
velocities in the thousands of km/s has been given in terms
of the frame-dragging effect (cf.~Fig.~5 of Ref.~\cite{Pretorius2007a}).

Investigating the momentum distribution and flow in recoiling 
BBH mergers could help to build further intuition into
the nonlinear dynamics of the spacetime and
their influence on the formation of kicks. 
Paper I made some headway into the former issue but could not
address the latter.
Specifically, paper I examined the distribution and the flow of linear 
momentum in BBH spacetimes  
using the Landau-Lifshitz formalism 
in the post-Newtonian approximation.
It 
then specialized this approach to the extreme-kick configuration
\cite{Campanelli2007a, Gonzalez2007b,Campanelli2007}, 
which is a system of
inspiraling, BBHs
with equal and anti-parallel spins in the 
orbital plane.
During inspiral, the two black holes simultaneously and sinusoidally bob
perpendicularly to the orbital plane; 
in paper I, this motion was first recognized as arising
from the combined effect of frame dragging and spin-curvature coupling 
and then was found to arise from the
exchange of momentum between the near-zone gravitational field and the
black holes.

Because paper I
analyzed the system at a post-Newtonian level,
its analysis could not be extended 
to merger and beyond.
Consequently,  
it was not possible to address how the nonlinear
dynamics in the pre-merger near zone 
transitions into the final 
behavior of the merged black hole.
This paper (paper II) lets us begin to address this 
transition as
we study momentum flow during the plunge, merger, and
ringdown of BBHs
in \emph{full numerical relativity}.
Our study allows us, for example, to
examine how accurately Pretorius's intuitive picture applies during 
the merger and ringdown of a recoiling BBH merger.

\subsection{Overview and summary}
\label{sec:intro_overview}
\begin{figure}
\includegraphics[width=0.9\hsize]{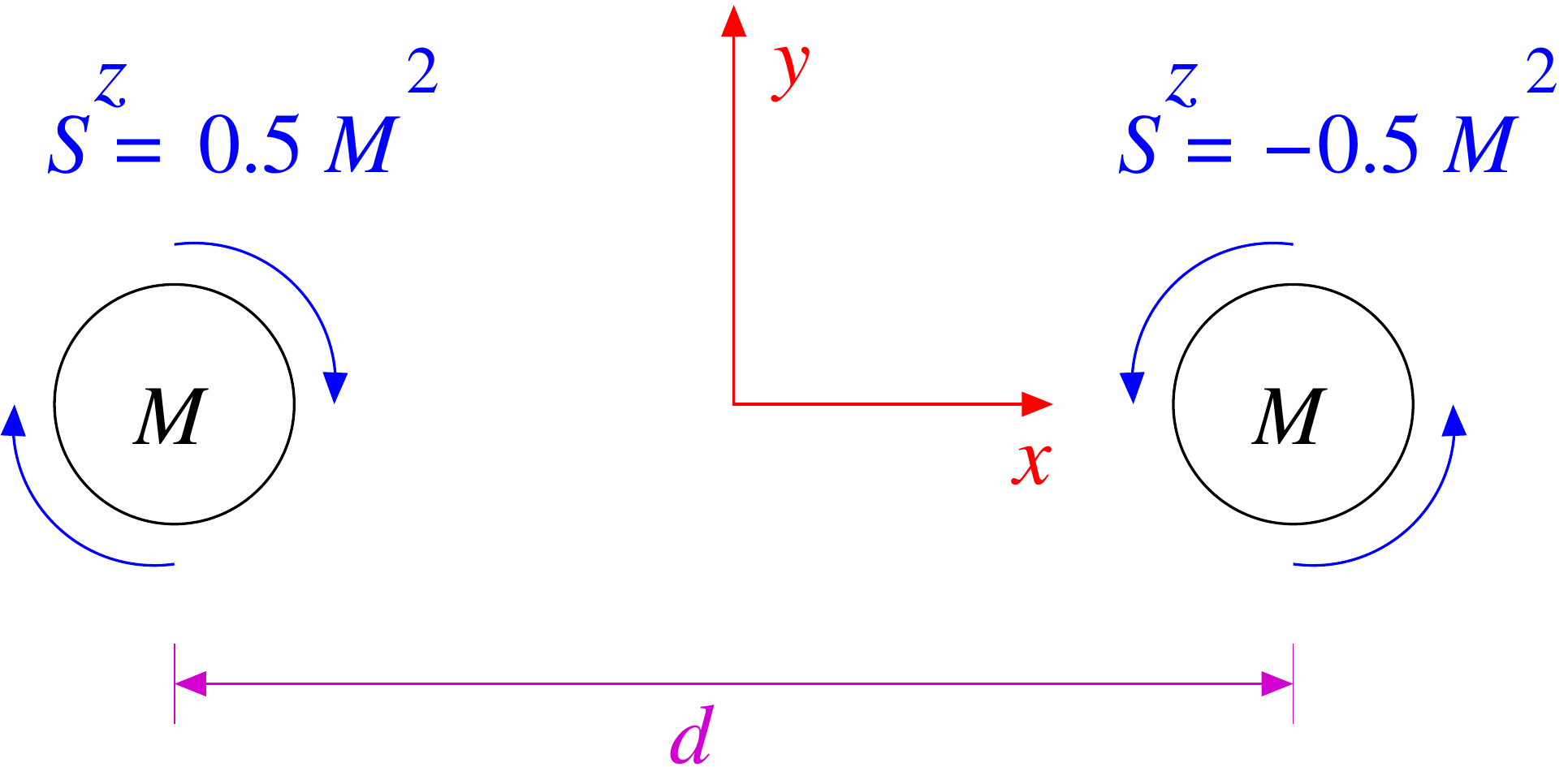}
\caption{Initial configuration of the 
head-on BBH considered in this paper. 
The holes move primarily along the $x$ axis, but they also 
accelerate in the $-y$ (downward) direction due to frame dragging. 
See Table~\ref{Table:SpECID} for the value of $d=2x_0$.
\label{fig:InitialConfiguration}}
\end{figure}

As a first step toward analyzing the momentum flow 
in superkicks, in this paper 
we apply the Landau-Lifshitz momentum-flow formalism to a much simpler 
case: 
the head-on plunge, merger, and ringdown of an equal-mass 
BBH. The holes initially have 
antiparallel spins of equal magnitude 
that are transverse to the holes' head-on motion
(Fig.~\ref{fig:InitialConfiguration}).
Primarily, the holes simply fall toward each other
in the $\pm x$ direction. 
However, each hole's spin drags the space around itself, 
causing the other hole to accelerate 
in the downward, $-y$ direction.  

How does this frame dragging relate to the final kick velocity of 
the merged hole? To address this question, we compute the 4-momentum 
$p^\mu$ 
inside each apparent horizon using the Landau-Lifshitz
formalism; we then define an \emph{effective velocity} as
\begin{eqnarray}\label{eq:veff_intro}
v^i_{\rm LL} := \frac{p^i}{p^0}.
\end{eqnarray}
In Sec.~\ref{sec:Results}, we find that this effective velocity behaves 
similarly to the horizons' coordinate velocities.

\begin{figure}
\includegraphics[width=3in]{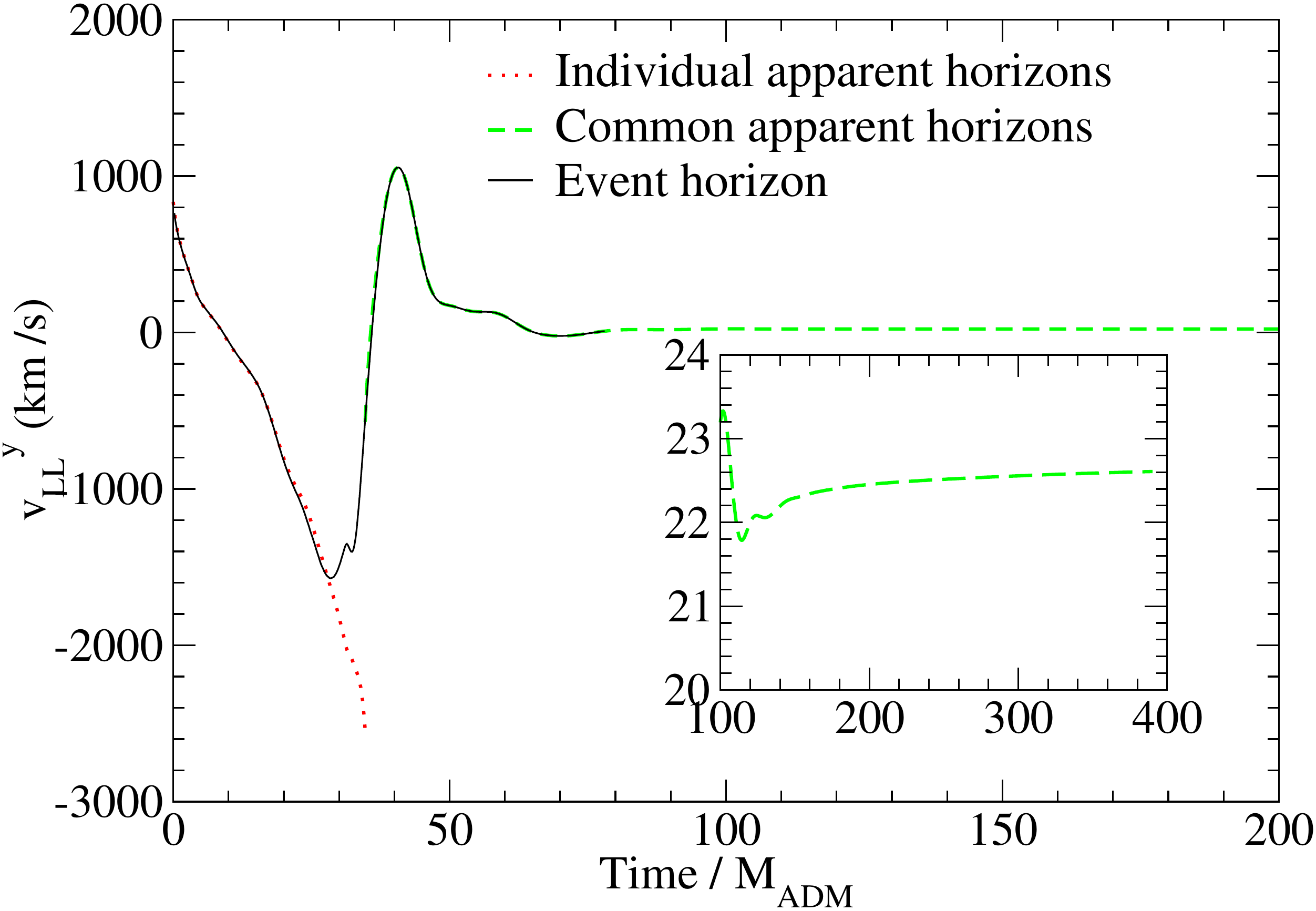} 
\caption{The effective velocity $v^y_{\rm LL}$ for the individual 
(red dotted line) and common (green dashed line) apparent horizons and 
for the event horizon (black solid line). 
The inset shows the velocity of the common apparent horizon at late times.
\label{fig:intro_results}}
\end{figure}

\begin{figure}
\includegraphics[width=3in]{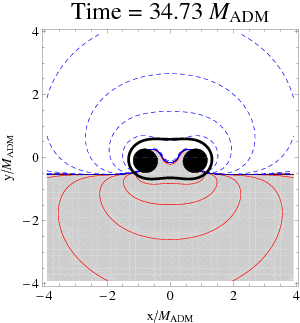}
\caption{A contour plot of the $y$ component of the momentum density 
at the moment when the common apparent horizon forms. The common horizon 
encloses the momentum inside the individual horizons and also the momentum 
in the gravitational field. The grey-shaded region and solid, red contours 
indicate 
positive momentum density, while the white-shaded region and blue, dashed 
contours indicate negative momentum density. The individual apparent horizons 
are shaded black, and the common apparent horizon is shown as a thick black 
line.
\label{fig:intro_contour}}
\end{figure}

The effective $y$ velocity for the spectral simulation described in 
Sec.~\ref{sec:EvolveSpEC} is shown in  
Fig.~\ref{fig:intro_results}. Before the merger, the individual 
apparent horizons do indeed accelerate in the $-y$ (``down'') 
direction, 
eventually 
reaching velocities of order $10^3$ km/s. However, 
when the common apparent horizon forms, it 
pulsates; during the first half-pulsation, the 
horizon expands and accelerates to $\sim 10^3$ km/s in the 
\emph{up} ($+y$) direction. This
happens
because as the common horizon 
forms and expands, 
it swallows not only the downward linear momentum 
inside each 
individual horizon but also a large amount of  
upward momentum in the 
gravitational field 
between the holes (Fig.~\ref{fig:intro_contour}). 
During the next half-pulsation,
as the horizon 
shape 
changes from oblate to
prolate (cf. Fig.~\ref{fig:Shape}), 
the horizon swallows a net downward momentum,
thereby 
losing most of its upward velocity.
Eventually, after strong damping of the pulsations,
the common horizon settles down to a 
very small 
velocity of about 23 km/s in the $+y$ direction
(inset of Fig.~\ref{fig:intro_results}), which 
(Sec.~\ref{sec:Results}) is consistent with the kick velocity inferred from the 
emitted gravitational radiation. 

This momentum flow between field and holes is also described quite 
beautifully in
the language of the holes' \emph{event horizon}.  
Unlike apparent horizons, 
the event horizon evolves and
expands continuously in time, rather than discontinuously.  
As the event horizon expands, it continuously swallows 
surrounding field
momentum, and
that swallowing produces a continuous evolution of 
the event horizon's velocity, an evolution
that is nearly the same as for the apparent-horizon velocity. 
Figure~\ref{fig:intro_results} shows how the effective velocity of the 
event horizon smoothly transitions from matching the individual 
apparent horizons' velocities to matching the common 
apparent horizon's velocity. For further details,
see Sec.~\ref{sec:EHresults} and especially Figs.~\ref{fig:Ev_3FC_Vy_JustEH} 
and 
\ref{fig:event_horizons}. 


In the remainder of this paper, we discuss our
results and the 
simulations that are used to obtain them.
In Sec.~\ref{sec:LLformal}, 
we briefly review the Landau-Lifshitz formalism and momentum conservation. 
The simulations themselves are 
presented in Sec.~\ref{sec:Simulations}. We analyze the simulations' 
momentum flow in Sec.~\ref{sec:Results} and 
conclude in 
Sec.~\ref{sec:Conclusions}. 
In the appendices, 
we describe in greater depth
the numerical methods used for the simulations presented in this paper.

\section{4-Momentum Conservation in the Landau-Lifshitz Formalism}
\label{sec:LLformal}

In this section, we briefly review the Landau-Lifshitz formulation of gravity 
and the statement of 4-momentum conservation within this theory.
Landau and Lifshitz, in their {\it Classical Theory of Fields} 
(hereafter referred to as LL),
reformulated general relativity as a nonlinear field theory in flat spacetime 
\cite{Landau-Lifshitz}.
(Chap. 20 of MTW \cite{MTW} and a paper by Babak and Grishchuk 
\cite{Babak-Grishchuk} 
are also helpful sources that describe the formalism.)
Landau and Lifshitz develop their formalism by first 
laying down arbitrary 
asymptotically Lorentz coordinates on a given 
curved (but asymptotically-flat) spacetime. 
They use these coordinates to map the 
curved (i.e. physical) spacetime onto an auxiliary flat spacetime 
(AFS) by enforcing that the 
coordinates on the AFS are
globally Lorentz. 
The auxiliary flat metric takes the Minkowski form, 
$\eta_{\mu\nu} = {\rm diag}(-1,1,1,1)$.

In this formulation, gravity is described 
by the physical metric density
\begin{equation}
\gg^{\mu \nu} := \sqrt{-g} g^{\mu \nu}\;,
\label{eq:gg}
\end{equation}
where $g$ is the determinant of the covariant components of the 
physical metric, and $g^{\mu\nu}$
are the contravariant components of the physical metric.
When one defines the superpotential
\begin{equation}
H^{\mu\alpha\nu\beta} := \gg^{\mu \nu}\gg^{\alpha \beta} 
- \gg^{\mu \alpha}\gg^{\nu \beta}\;,
\label{eq:superpotential}
\end{equation}
the Einstein field equations take the field-theory-in-flat-spacetime form
\begin{equation}
{H^{\mu\alpha\nu\beta}}_{,\alpha\beta} = 16\pi \tau^{\mu\nu}\;.
\label{eq:efe}
\end{equation}
Here $\tau^{\mu\nu} := (-g)(T^{\mu\nu} + t^{\mu\nu}_{\rm LL})$ is the 
total effective stress-energy tensor,
indices after the comma denote partial derivatives 
or, equivalently, covariant derivatives 
with respect to the flat
auxiliary metric), and the Landau-Lifshitz pseudotensor $t^{\mu\nu}_{\rm LL}$ 
(a real tensor in the
auxiliary flat spacetime) is given by Eq.\ (100.7) of 
LL \cite{Landau-Lifshitz} or equivalently Eq.\ (20.22) of MTW
\cite{MTW}:
\begin{eqnarray}\label{eq:LLPseudo}
16\pi (-g) t^{\alpha\beta}_{\rm LL} & = & 
\gg^{\alpha\beta}{_{,\lambda}}\gg^{\lambda\mu}{_{,\mu}}
-\gg^{\alpha\lambda}{_{,\lambda}}\gg^{\beta\mu}{_{,\mu}} \nonumber\\
& + & \frac{1}{2}g^{\alpha\beta}g_{\lambda\mu}
\gg^{\lambda\nu}{_{,\rho}}\gg^{\rho\mu}{_{,\nu}} \nonumber\\
& - & g^{\alpha\lambda}g_{\mu\nu}\gg^{\beta\nu}{_{,\rho}}\gg^{\mu\rho}{_{,\lambda}}
-g^{\beta\lambda}g_{\mu\nu}\gg^{\alpha\nu}{_{,\rho}}\gg^{\mu\rho}{_{,\lambda}}
\nonumber\\
& + & g_{\lambda\mu}g^{\nu\rho}\gg^{\alpha\lambda}{_{,\nu}}\gg^{\beta\mu}{_{,\rho}}
\nonumber\\
& + & \frac{1}{8}
\left(2g^{\alpha\lambda}g^{\beta\mu}-g^{\alpha\beta}g^{\lambda\mu}\right)
\nonumber\\ & \times &
\left(2g_{\nu\rho}g_{\sigma\tau}-g_{\rho\sigma}g_{\nu\tau}\right)
\gg^{\nu\tau}{_{,\lambda}}\gg^{\rho\sigma}{_{,\mu}}
\end{eqnarray}
Due to the symmetries of the superpotential---they are the same as those 
of the Riemann tensor---the field equations 
(\ref{eq:efe}) imply the differential conservation law for 4-momentum
\begin{equation}
{\tau^{\mu\nu}}_{,\nu} = 0\;.
\label{eq:divtau}
\end{equation}
Eq.\ (\ref{eq:divtau}) is equivalent to ${T^{\mu\nu}}_{;\nu} = 0$, where the 
semicolon denotes a covariant
derivative with respect to the physical metric.

In both LL and MTW, it
is shown that the total 4-momentum of any isolated 
system (measured in the asymptotically 
flat region far from the system) is
\begin{equation}
p^\mu_{\rm tot} = \frac{1}{16\pi}\oint_{\mathcal S} {H^{\mu\alpha 0 j}}_{,\alpha} 
d\Sigma_j\;,
\label{eq:MomTot}
\end{equation}
where $d\Sigma_j$ is the surface-area element of the flat auxiliary metric, 
and $\mathcal S$ is an arbitrarily large
surface surrounding the system.  
This total 4-momentum satisfies the usual conservation law
\begin{equation}
\frac{dp^\mu_{\rm tot}}{dt} = - \oint_{\mathcal S} \tau^{\mu j} d\Sigma_j\;.
\label{eq:dMomTotdt}
\end{equation}
See the end of Section III of \cite{Chen2009} for a brief proof of why this 
holds for black holes. 

\begin{figure}
\includegraphics[width=0.6\columnwidth]{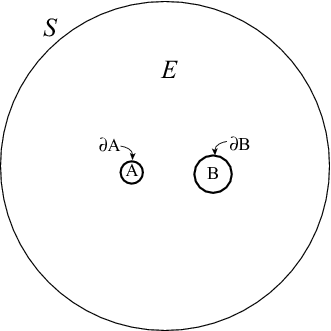}
\caption{The regions of space around and inside a binary-black-hole system.}
\label{fig:Binary1}
\end{figure}

Because this paper focuses on BBHs, we will make a few further 
definitions that will be used
frequently in our study.
First, we label the two\footnote{After the holes merge, 
there is only one horizon, which we 
label $\partial C$. 
Equations
(\ref{eq:dMomTotdt})--(\ref{eq:veff}) hold after removing terms with 
subscript $B$ and then substituting 
$A\rightarrow C$.}
black holes in the binary 
(and the regions of space within their horizons)
by $A$ and $B$, and denote their surfaces 
(sometimes the hole's event horizon and other times the apparent 
horizon) by $\partial A$ and $\partial B$, as shown in 
Fig.~\ref{fig:Binary1}.  
We let $\mathcal E$ stand for the region outside both bodies 
but inside the arbitrarily large surface
$\mathcal S$ where the system's total momentum is computed 
(in our case, this is taken to be 
a fixed coordinate sphere inside 
the outer boundary of the numerical-relativity 
computational grid).

With the aid of Gauss's theorem and the Einstein field equations 
(\ref{eq:efe}), one can reexpress
Eq.\ (\ref{eq:MomTot}) for the binary's total 4-momentum as a sum over 
contributions from each of the bodies
and from the gravitational field in the region $\mathcal E$ outside them:
\begin{subequations}
\begin{equation}
p_{\rm tot}^\mu  = p_A^\mu + p_B^\mu + p_{\rm field}^\mu\;.
\label{eq:pDecompose}
\end{equation}
Here
\begin{equation}
p_A^\mu := \frac{1}{16\pi}\oint_{\partial A} {H^{\mu\alpha 0 j}}_{,\alpha} 
d\Sigma_j
\label{eq:pAsurf}
\end{equation}
is the 4-momentum of body $A$ (an equivalent expression holds for body $B$),
and
\begin{equation}
p_{\rm field}^\mu := \int_{\mathcal E} \tau^{0\mu} d^3 x
\label{eq:pEvol}
\end{equation}
is the gravitational field's 4-momentum in the exterior of the black holes.
\label{eq:pComponents}
\end{subequations}
We 
define an effective velocity of black hole $A$ (with 
similar expressions holding for hole $B$) by
\begin{equation}
v_{\rm LL}^j := \frac{p^j_A}{p^0_A}\;.
\label{eq:veff}
\end{equation}

In analogy to Eq.\ (\ref{eq:dMomTotdt}) for the rate of change of the 
binary's total 4-momentum, 
one can write the corresponding equation for the rate of change of the 
4-momentum of body $A$:
\begin{equation}
\frac{dp^\mu_A}{dt} = - \oint_{\partial A} (\tau^{\mu k} - \tau^{\mu 0} v^k_A ) 
d\Sigma_k\;.
\label{eq:dpAdtsurf}
\end{equation}
Equation (\ref{eq:dpAdtsurf}) describes the flow of field 4-momentum into and 
out of body $A$
(the second term comes from the motion of the boundary of body $A$ with 
local coordinate velocity $v^k_A$).\footnote{In the case that the 
body's event horizon is stationary (i.e. sufficiently
 far from merger), $v^k_A = dx_{A\;{\rm cm}}^k/dt$, the center of 
mass velocity of body $A$.  However, if the body's event horizon is 
dynamical (i.e. during the merger phase), then $v^k_A$ is the local coordinate
 velocity of the event horizon surface, $v^k_A = dx_{\partial A}^k/dt$. 
See Sec.~\ref{sec:EHresults} for a discussion of the dynamics of the
event horizon. }

We will use Eqs.\ (\ref{eq:dMomTotdt})--(\ref{eq:veff})
as the basis for our study of momentum flow in black-hole binaries.
The actual values of the body and field 4-momenta, computed in the above ways, 
will depend on the 
arbitrary 
mapping between the physical spacetime and 
the AFS;
this is the gauge-dependence that will be 
discussed in 
Sec.~\ref{sec:Gauge}.

\section{Simulations of head-on BBH collisions with 
anti-aligned spins}
\label{sec:Simulations}

In order to investigate the gauge dependence of our results, we
compare simulations of the same physical system using two separate
methods that employ different choices of coordinates.  One method is a
pseudospectral excision scheme based on generalized harmonic
coordinates; the other is a finite-difference moving-puncture scheme
that uses 1+log slicing and a gamma-driver shift condition (henceforth
referred to as ``moving puncture gauge''; for details see
Appendix~\ref{app: movpunc}). The coordinates used in the two methods
differ both for the initial data and during the evolution.  In this
section we summarize the construction of initial data and the
evolution scheme for both methods, and we present convergence tests
and estimate numerical uncertainties.  Further details about our
numerical methods are are given in Appendices~\ref{sec:AppID}
and~\ref{sec:numer-meth-evol}.

\subsection{Pseudospectral}

\subsubsection{Quasiequilibrium excision data}\label{sec:IDSpEC}
\label{sec:SpECID}
The evolutions described in Sec.~\ref{sec:EvolveSpEC} begin with 
quasiequilibrium excision data constructed using the method of 
Ref.~\cite{Lovelace2008}. 
This method requires the arbitrary choice of a conformal
three-metric; we choose this metric to be flat 
almost everywhere
but curved (such that 
the metric is nearly that of a single Kerr-Schild hole) near the horizons. 

Our initial data method
also requires us to choose an outer boundary condition
on a shift vector $\beta^i$; 
for a general binary that is orbiting
and inspiraling,
we use\footnote{
The shift vector $\beta^i$ used here and in 
Appendix~\ref{sec:AppID} for the construction 
of initial data is not the same as the shift vector used during our 
evolutions. Except for Sec.~\ref{sec:SpECID} and Appendix~\ref{sec:AppID}, 
we always use $\beta^i$ to refer to the shift \emph{during the evolution}.}
\begin{eqnarray}
\beta^i = \left(\mathbf \Omega_0 \times r\right)^i + \dot{a}_0 r^i 
+ V_0^i, \mbox{ } r\rightarrow\infty ,
\label{eq:shiftbc}
\end{eqnarray} 
where $\Omega_0$ is the angular velocity, $\dot{a}_0 r^i$ is the
initial radial velocity, and $V_0^i$ is a translational velocity.
Note that Eq~(\ref{eq:shiftbc}) is different from the choice
made in Ref.~\cite{Lovelace2008}.  In this paper we confine our focus
to collisions that are head-on, which we define as $\Omega_0=\dot{a}_0=0$.
However, $V_0^i$ must be 
nonzero to make 
the total linear momentum of the initial data vanish. 

Table~\ref{Table:SpECID} summarizes the 
initial data used in this paper. 
The Arnowitt-Deser-Misner (ADM) mass
$M_{\rm ADM}$ (Eq.~(11.2.14) in Ref.~\cite{Wald}; see also
\cite{ADM,york79}), 
the irreducible mass $M_{\rm irr}$ and  
Christodoulou 
mass $M_{\rm Chr}$ of one of the holes are listed, where $M_{\rm Chr}$ is 
related to $M_{\rm irr}$ and the spin of the hole $S_z$ by
\begin{eqnarray}
M_{\rm Chr}^2 = M_{\rm irr}^2 + \frac{S_z^2}{4 M_{\rm irr}^2}.
\end{eqnarray} 
Table~\ref{Table:SpECID} also shows the dimensionless spin 
$S_z/M_{\rm Chr}^2$; by definition, this measure of the spin lies in the 
interval $-1 \leq S_z/M_{\rm Chr}^2 \leq 1$.

For set S1 listed in 
Table~\ref{Table:SpECID},
$V_0^i$ is adjusted so that the initial 
effective velocity of the entire spacetime 
$v^i_{\rm tot}:=p^i_{\rm tot}/p^0_{\rm tot}$ 
is smaller than 0.1 km/s, 
which is approximately the size 
of our numerical truncation error (cf. Fig.~\ref{fig:vLLDiffVsResolution}):
$\left(\left|v^x_{\rm tot}\right|,\left|v^y_{\rm tot}\right|,
\left|v^z_{\rm tot}\right|\right) 
= (4\times 10^{-4},5\times 10^{-2},2\times10^{-3})$ km/s at time $t=0$.

The construction
of initial data is described in 
more detail in Appendix~\ref{sec:AppID}.
\begin{table}
\begin{ruledtabular}
\begin{tabular}{c|cccc}
Set & $x_o/M_{\rm ADM}$ & $M_{\rm irr}/M_{\rm ADM}$ 
& $M_{\rm Chr}/M_{\rm ADM}$ 
& $S_z / M_{\rm Chr}^2$\\
\hline
S1 &  $3.902$   &  0.4986  &  0.5162  &  $\pm0.5000$ \\
\hline
P1 &  $4.211$  &  0.4970  &  0.5146  &  $\pm 0.5000$  \\
P2 &  $8.368$  &  0.4802  &  0.5072  &  $\pm 0.5091$  \\
\hline
H1 &  $14.864$ &  0.4870  &  0.5042  &  $\pm 0.5000$ \\
\end{tabular}
\end{ruledtabular}
\caption{\label{Table:SpECID} \label{tab: lean_models}
         Parameters of the initial data configurations studied 
         in this work.
         Model S1 (see Sec.~\ref{sec:SpECID})
         gives the parameters used to construct 
         a set of Superposed-Kerr-Schild quasiequilibrium excision 
         initial data.  Model H1 (see Appendix~\ref{sec:AppSHK}) 
         gives the parameters for the larger 
         separation Superposed-Harmonic-Kerr initial data set.  Both S1
         and H1 were used in spectral evolutions.  P1 and P2 provide 
         the Bowen-York parameters for the two systems evolved with the 
         moving puncture method.  
         The holes are initially 
         separated by a coordinate distance $d=2 x_0$ and are located
         at coordinates $(x,y,z)=(\pm x_0,0,0)$.
         For clarity, only 
         4 significant figures are shown.
        }
\end{table}

\subsubsection{Generalized harmonic evolutions}\label{sec:EvolveSpEC}

We evolve the quasiequilibrium excision data described in 
Sec.~\ref{sec:SpECID} pseudospectrally, using generalized harmonic 
gauge~\cite{Friedrich1985,Garfinkle2002,Pretorius2005c,Lindblom2006}, 
for which the coordinates $x^\mu$ satisfy the gauge condition
\begin{eqnarray}\label{eq:SpECGauge}
g_{\mu\nu} \nabla^{\rho}\nabla_{\rho} x^\mu = H_\nu \left(x^\rho, g_{\sigma\tau} 
\right)
\end{eqnarray} where $H_\nu$ is a function of the coordinates and the spacetime 
metric.
In this subsection, 
we summarize the computational grid used for our spectral evolutions, 
and we briefly discuss our numerical accuracy.
Details of our pseudospectral evolutions are given in 
Appendix~\ref{sec:pseud-evol}.

Our computational grid covers only the exterior regions of the black
holes (``black hole excision''): there is an artificial inner boundary
just inside each apparent horizon where no boundary condition is needed
because of causality.  The grid extends to a large radius 
$r_{\rm max}\sim 400 M_{\rm ADM}$.  
A set of overlapping subdomains of different
shapes (spherical shells near each hole and far away; cylinders elsewhere)
covers the entire space between the excision boundaries and $r=r_{\rm max}$.

Because different subdomains have different shapes and the grid
points are not distributed uniformly, we describe the
resolution of our grid in terms of the total number of grid points summed
over all subdomains.  We
label our resolutions $\mbox{N}0$, $\mbox{N}1$, and $\mbox{N}2$, corresponding
to approximately 
$55^3$, $67^3$, and $79^3$ grid points, respectively.
After merger, we regrid onto a new computational domain that has only
a single excised region (just inside the newly-formed apparent horizon
that encompasses both holes). This new grid has a different resolution
(and a different decomposition into subdomains) from the old grid.
We label the resolution of the post-merger grid by $A$, $B$, and $C$,
corresponding to approximately $63^3$, $75^3$, 
and $87^3$ gridpoints, respectively.  We label the entire run using the
notation `N$x$.$y$', where the characters before and after the
decimal point denote the pre-merger and post-merger resolution
for that run. Thus, for example, `$\mbox{N}2.B$' denotes
a run with approximately $67^3$ grid points before merger, and
$75^3$ grid points afterward.
On the outermost portion of the grid (farther than 
$\sim 200 M_{\rm ADM}$), 
we use a coarser numerical 
resolution than we do elsewhere.
(We only measure the gravitational wave flux, linear momentum, 
etc., at radii of $r\le 160 M_{\rm ADM}$.)

\begin{figure}
\includegraphics[width=3.5in]{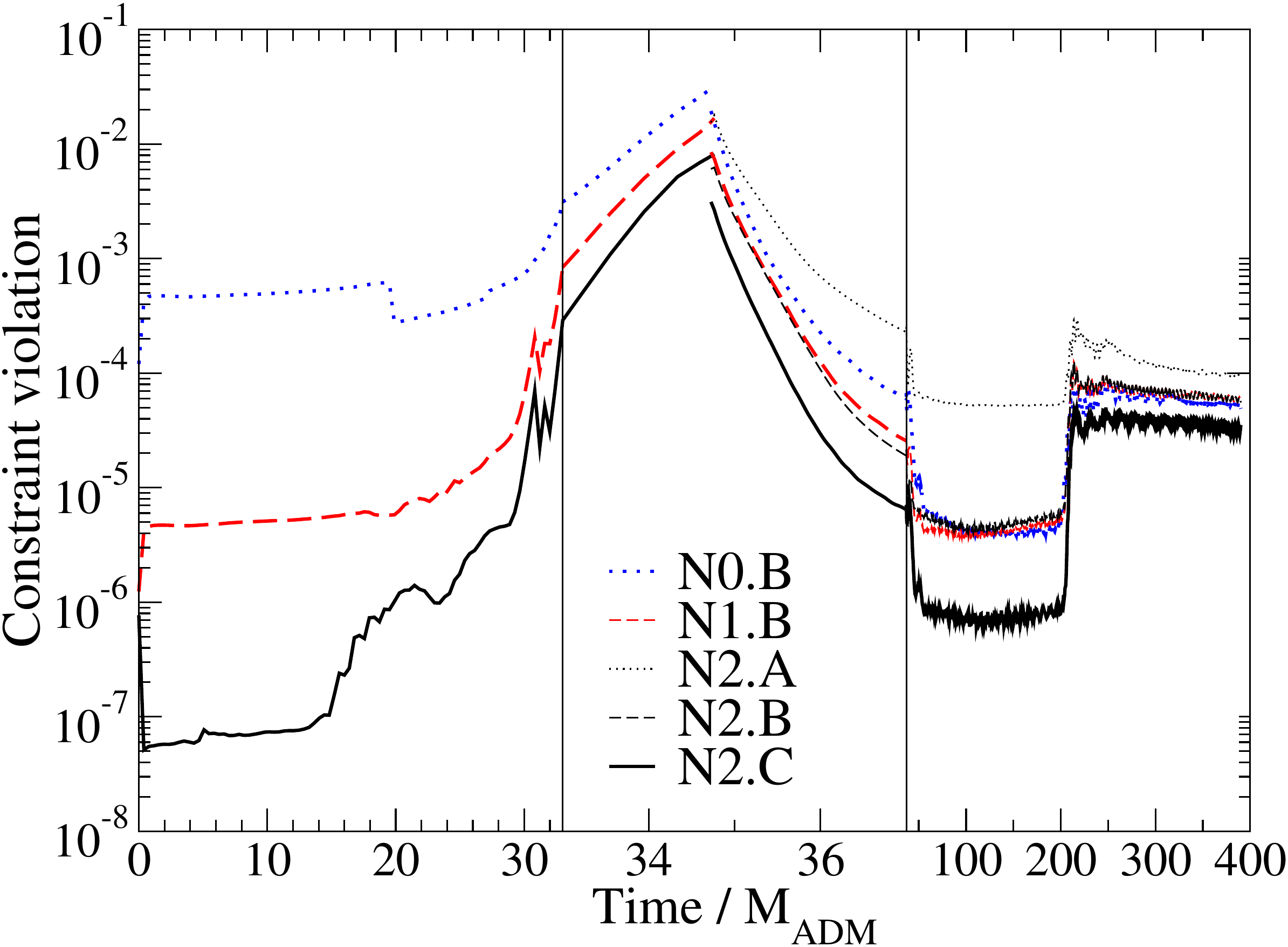}
\caption{Constraint violation at different numerical
resolutions for the 
pseudospectral evolutions S1. The common apparent horizon forms at time 
$t=34.73 M_{\rm ADM}$. 
Labels of the form $\mbox{N}x.y$ indicate the grid resolution, where
the pre-merger resolution is labeled (from coarse to fine) by $x=0,1,2$ 
and the post-merger resolution is labeled by $y=A,B,C$.
The constraints decrease exponentially with higher resolution; the convergence
rate is smaller near merger.
\label{fig:Ev_3FC_NormalizedGhCe}
}
\end{figure}

To demonstrate the convergence of our evolutions, we plot the constraint 
violation in Fig.~\ref{fig:Ev_3FC_NormalizedGhCe} for several
resolutions. The quantity plotted is the $L^2$ norm of all the
constraints of the generalized harmonic system, normalized by the
$L^2$ norm of the spatial gradients of all the dynamical fields, as defined
by by Eq.~(71) of Ref.~\cite{Lindblom2006}.
The left portion of the 
plot depicts 
the constraint violation during 
the plunge, the right third of the 
plot shows the constraint violation during the ringdown, and the 
middle panel shows the constraints shortly before and shortly after the common 
apparent horizon forms. Throughout the evolution, we generally observe 
exponential convergence, although
the convergence rate is smaller near merger.
After merger, there are two sources of constraint violations: those
generated by numerical truncation error after merger (these depend on the
resolution of the post-merger grid) and those generated by numerical
truncation error before merger and are still present in the solution
(these depend on the resolution of the pre-merger grid).
We see from Fig.~\ref{fig:Ev_3FC_NormalizedGhCe} that the constraint violations
after merger are dominated by the former source.
Also, at about $t=200 M_{\rm ADM}$, 
the constraint violation increases noticeably 
(but is still convergent);
at this time, the outgoing gravitational waves have reached 
the coarser, outermost 
region of the grid.

\begin{figure}
\includegraphics[width=3.5in]{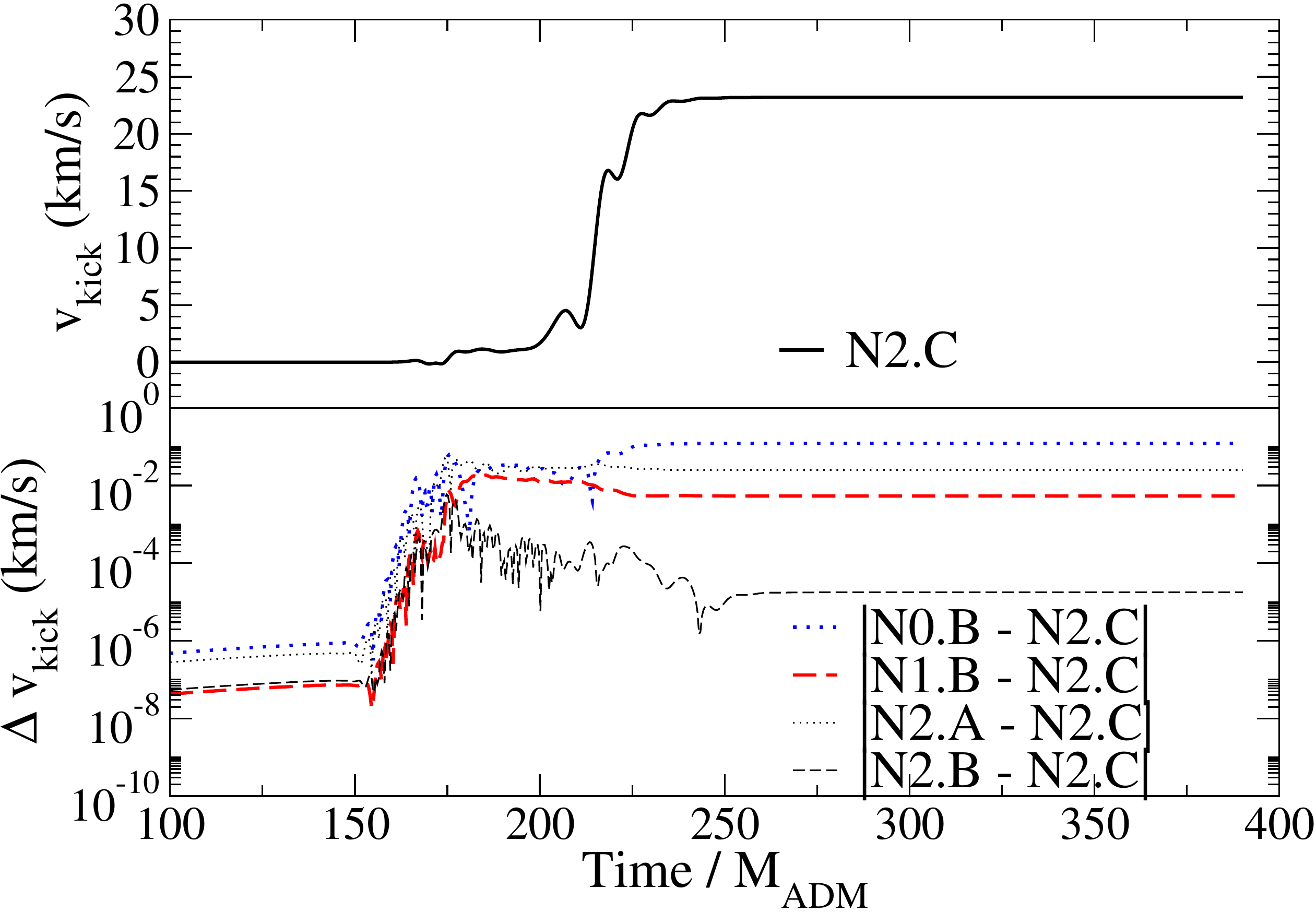}
\caption{Recoil velocity for 
initial data set S1 inferred 
from the gravitational wave signal 
$\Psi_4$ extracted at $r_{\rm extr}=160 M_{\rm ADM}$ at the highest 
resolution (upper panel). Differences between several coarser resolutions 
and the highest resolution result are plotted in the lower panel.
\label{fig:S1kickPsi4}}
\end{figure}

Finally, in Fig.~\ref{fig:S1kickPsi4}, we demonstrate the accuracy of the 
recoil velocity $v_{\rm kick}=22$ km/s
inferred from the gravitational wave signal 
$\Psi_4$, which asymptotically is related to the gravitational wave 
amplitudes $h_+$ and $h_\times$ by
\begin{eqnarray}
\Psi_4 = \frac{d^2}{dt^2} h_{+} - i \frac{d^2}{dt^2} h_{\times}.
\label{eq:psi4}
\end{eqnarray}
We extract the spin-weighted spherical harmonic coefficients
of $\Psi_4(t)$ 
from the simulation as described in Ref.~\cite{Scheel2008}, and we
integrate these coefficients over time to obtain $\dot{h}^{\ell m}(t)$, which
are the spin-weighted spherical harmonic
coefficients of $\dot{h}= \dot{h}_{+} - i \dot{h}_{\times}$. 
For each 
$(\ell,m)$, the integration
constant is chosen so that the average value of $\dot{h}^{\ell m}(t)$ is zero.
The $\dot{h}^{\ell m}(t)$ are then used to compute the
4-momentum flux of the gravitational waves from
Eqs.~(3.14)--(3.19) of
Ref.~\cite{RuizTakAlcNunez2007}. Integrating this flux over
time yields the total radiated energy-momentum, $p^\mu_{\rm rad}$. The recoil
velocity can then be computed from energy-momentum conservation:
$v^i_{\rm kick}=-p^i_{\rm rad}/M_{\rm final}$, 
where $M_{\rm final} := M_{\rm ADM} - E_{\rm rad}$ and $E_{\rm rad}$ is the 
energy radiated to infinity. For set S1, we obtain a radiated energy of 
$E_{\rm rad}/M_{\rm ADM}=(5.6840 \pm 0.0008)\times 10^{-4}$, where the 
quoted error includes truncation error and uncertainty from extrapolation 
to infinite radius (as discussed below).
The top panel of Fig.~\ref{fig:S1kickPsi4}
shows the recoil velocity as a function of time for our highest resolution 
simulation, while the lower panel shows differences between the 
highest resolution ($\mbox{N}2.C$) and 
lower resolutions. 
From these differences, we  
estimate a numerical uncertainty for the final
recoil 
velocity of $5\times 10^{-3}$ km/s for $\mbox{N}1.B$ 
and $2\times 10^{-5}$ km/s for 
$\mbox{N}2.B$. 

This numerical uncertainty includes only the effects of
numerical truncation error; however, there are other potential
sources of uncertainty in the simulations that 
must also be considered.
The first is the spurious ``junk'' gravitational radiation that
arises because the initial data do not describe a perfect equilibrium
situation.  This radiation 
is not astrophysically realistic, but by carrying
a small amount of
energy-momentum that contributes to the measured $p^\mu_{\rm rad}$
at large distances, the spurious radiation does affect 
our determination of the final recoil
velocity.
In our investigation of momentum flow (Sec.~\ref{sec:Results}), 
we do not correct for the initial data's 
failure to be in equilibrium; here we estimate the contribution of 
the resulting spurious radiation to the final recoil velocity.
First, we note that for head-on collisions, the physical 
gravitational waves are emitted
predominantly after merger. Therefore, 
we estimate the influence of the spurious radiation by 
examining the accumulated recoil velocity at time 
$t=\Delta t+r$, where $r$ is the radius of the extraction 
surface and $\Delta t$ is a cutoff time. Because the holes merge so quickly 
(because they begin at so small an initial separation), 
the spurious and physical contributions to the recoil are not clearly 
distinguishable in Fig.~\ref{fig:S1kickPsi4}. Varying $\Delta t$ between
$31.1 M_{\rm ADM}$ and $38.3 M_{\rm ADM}$ (the common event and apparent horizons 
form at $t=31.1 M_{\rm ADM}$ and $t=34.7 M_{\rm ADM}$, respectively),
we estimate that the 
spurious radiation contributes approximately $1$ km/s (about 5\%) 
to the recoil 
velocity---a much larger uncertainty than the truncation error. 
The same variation of $\Delta t$ implies that the spurious radiation 
contributes 
about 10\% of the total radiated energy $E_{\rm rad}$.)

Another potential source of uncertainty in $v^i_{\rm kick}$ arises from
where on the grid we measure the gravitational radiation.  
In particular, the quantity
$\Psi_4$ in Eq.~(\ref{eq:psi4}) should ideally be measured at future
null infinity.  Instead, we measure $\Psi_4$ on a set of coordinate spheres
at fixed radii, compute $v^i_{\rm kick}$ on each of these spheres, and
extrapolate the final equilibrium value of $v^i_{\rm kick}$ to infinite
radius.  The dotted curves on Fig.~\ref{fig:SpECvelFinal} show
$v^y_{\rm kick}$ measured from $\Psi_4$ at several radii, and the black
cross shows the final value of $v^y_{\rm kick}$ extrapolated to infinity.
We estimate our uncertainty in the extrapolated value by comparing
polynomial extrapolation of orders $1$, $2$, and $3$; we find an
uncertainty of $3\times 10^{-3}$ km/s for the quadratic fit.
Note that if we had not extrapolated to infinity, but had instead
simply used the value of $v^y_{\rm kick}$ at our largest extraction
sphere ($r=160 M_{\rm ADM}$), we would have made an error of $0.85$
km/s, which is much larger than the uncertainty from numerical truncation error.
Finally, we mention that our computation of $\Psi_4$ is not strictly gauge
invariant unless $\Psi_4$ is evaluated at future null infinity.  As long
as gauge effects in $\Psi_4$ fall off faster than $1/r$ as expected,
extrapolation of
$v^y_{\rm kick}$ to infinity should eliminate this source of uncertainty.

\subsection{Moving puncture}
\subsubsection{Bowen-York puncture data}
\label{sec: inidata_lean}
In order to address the importance of 
gauge dependence for our calculations using the 
Landau-Lifshitz formalism, we also 
simulate BBH mergers using the so-called moving puncture method, which employs
the covariant form of ``1+log'' slicing \cite{Campanelli2006a, Bona1997}
for the lapse function $\alpha$ and a ``Gamma-driver''
condition (based on the original ``Gamma-freezing'' condition introduced in
\cite{Alcubierre2002}) for the shift vector. The precise
evolution equations for the gauge variables as well as further technical
details of our puncture simulations are given in Appendix~\ref{app: movpunc}.

Our simulations start with puncture initial data
\cite{Brandt1997} provided in our case by the spectral solver
of Ref.~\cite{AnsorgBruegmann2004}. The initial data are fully
specified in terms of the initial spin $\vec{S}_{1,2}$,
linear momentum $\vec{P}_{1,2}$ and initial coordinate
position $\vec{x}_{1,2}$ as well as the bare mass
parameters $m_{1,2}$ of either hole \cite{Bowen-York:1980}.
The corresponding nonvanishing values for the two puncture
models considered in this work are given in Table \ref{tab: lean_models}.
There we also list the total black-hole mass $M_{\rm Chr}$ and normalize
all quantities using the total ADM mass $M_{\rm ADM}$.
The main difference
between the two configurations is the initial separation of the holes.
The lapse and shift are initialized as 
$\alpha = \gamma^{-1/6}$ and $\beta^i=0$,
where $\gamma$ is the determinant of the physical three-metric.

\subsubsection{Moving puncture evolutions}\label{sec:punctureEvolve}
The evolution of the puncture initial data is performed using
sixth order spatial discretization of the BSSN equations combined
with a fourth order Runge-Kutta time integration. Mesh refinement
of Berger-Oliger \cite{Berger1984}
type is implemented using Schnetter's {\sc Carpet}
package \cite{SchnetterHawleyHawke2004, Carpetweb}.
The prolongation operator is of fifth order in space and quadratic in time.
Outgoing radiation boundary conditions are implemented using
second-order accurate advection derivatives (see, for example, Sec.~VI in
Ref.~\cite{Alcubierre2003a}).
\begin{figure}
\includegraphics[angle=0,width=3.0in]{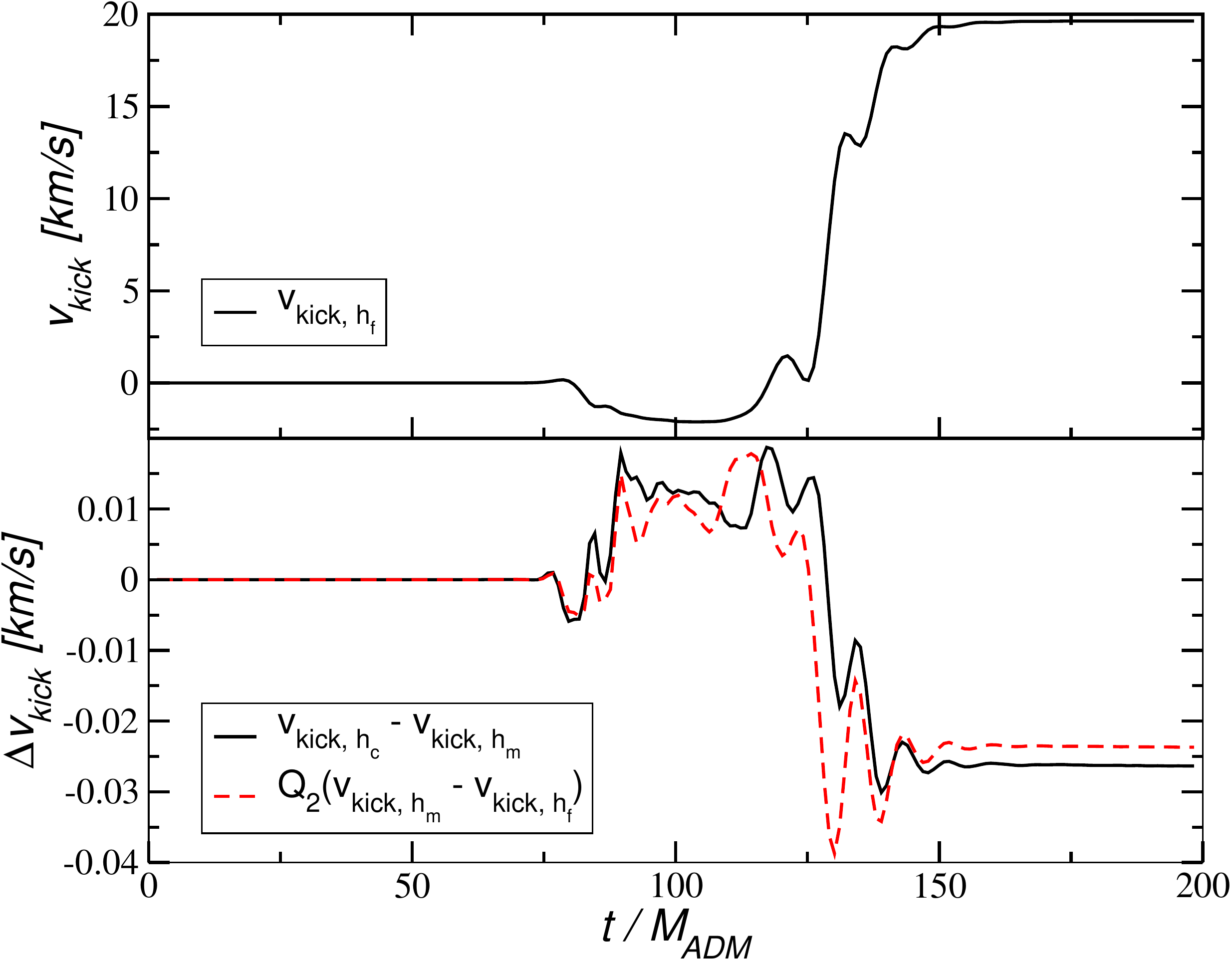}
\caption{Gravitational recoil for model P1
         as estimated from the gravitational wave signal
         $\Psi_4$ extracted at $r_{\rm ex}=73.5~M_{\rm ADM}$ using
         the highest resolution (upper panel).
         Differences in the recoil obtained at coarse, medium and
         fine resolution rescaled for second order convergence
         (lower panel).
        }
\label{fig: lean_conv}
\end{figure}

Using the notation of Sec.~II E of Ref.~\cite{Sperhake2006} the grid setup
in units of $M_{\rm ADM}$ for these evolutions is given by (rounded to 
3 significant digits)
\begin{eqnarray}
  \{(202,101,58.8,25.2,12.6) \times (3.15, 1.58,0.788),~h\}, \nonumber \\
  \{(201,100,58.5,25.1) \times (6.27,3.13,1.57,0.784),~h\}, \nonumber
\end{eqnarray}
respectively. Here $h$ denotes the resolution on the innermost refinement
level. For model P1 we perform a convergence analysis by setting $h$
to $h_{\rm c}=M_{\rm ADM}/49.5$, $h_{\rm m}=M_{\rm ADM}/57.1$
and $h_{\rm f}=M_{\rm ADM}/64.7$, respectively, for coarse, medium
and fine resolution. Model P2 is evolved using $h=M_{\rm ADM}/49.8$.

Before we discuss the physical results from the puncture simulations,
we estimate the numerical errors due to discretization, finite extraction
radius and the presence of unphysical gravitational radiation in the
initial data. 

In order to study the dependence of the results on resolution, we
have evolved model P1 of Table \ref{tab: lean_models} using
different resolutions $h_{\rm c}$, $h_{\rm m}$ and $h_{\rm f}$ on the
finest level and correspondingly larger grid spacings by a factor
of two on each consecutive level. The 
kick velocity
from the high resolution simulation, as inferred from the gravitational
radiation flux at $r_{\rm ex}=73.5~M_{\rm ADM}$,
is shown in the upper panel of Fig.~\ref{fig: lean_conv}. The bottom
panel shows the differences between the 
velocities
obtained at the different resolutions scaled for second order convergence
using a factor $Q_2=1.49$. By using Richardson extrapolation
we estimate the error in the final kick for the fine resolution
run to be $1~{\rm km/s}$ or $5~{\rm \%}$. We similarly find
overall second order convergence for the velocity derived
from the components of the Landau-Lifshitz tensor as integrated
over the apparent horizon. The error in that quantity barely varies
throughout the entire simulation and stays at a level just below
$\Delta v_{\rm LL} \approx 50~{\rm km/s}$ and $60~{\rm km/s}$
for fine and coarse resolution respectively.

The gravitational wave signal is further affected by the use of
finite extraction radius and linear momentum contained in the
spurious initial radiation.
We estimate the uncertainty due to the finite extraction radius by fitting
the final kick velocity obtained for the medium
resolution simulation of model P1 at radii $r_{\rm ex}=31.5...94.5~M_{\rm ADM}$
in steps of $10.5~M_{\rm ADM}$.
The resulting final kick velocities are well approximated by a polynomial
of the form $a_0 + a_1/r_{\rm ex} + a_2/r^2_{\rm ex}$.
For $r_{\rm ex}=73.5~M$ we thus obtain an uncertainty of $0.4~{\rm km/s}$
corresponding to a relative error of $2.2~\%$.

Finally we take into account contributions from the spurious initial
radiation by discarding the wave signal up to $t-r_{\rm ex} = \Delta t$.
For model P1 it is not entirely clear where exactly the spurious wave signal
stops and the physical signal starts. By varying $\Delta t$ from
$30$ to $45~M_{\rm ADM}$ we obtain an additional
error of about $\pm1~{\rm km/s}$.
For model P2 no such problem arises because of the smaller amplitude of the
spurious radiation and because the longer pre-merger time enables the
junk radiation to escape the system long before the merger happens.
We estimate the resulting total uncertainty by summing the squares of
the individual errors and obtain $7.5~\%$ and $5.5~\%$ for
models P1 and P2, respectively.

Using these uncertainties,
the gravitational wave emission for model P1 results in a
total radiated energy of $E_{\rm rad}/M_{\rm ADM}=(0.042 \pm 0.008)~\%$
and a recoil velocity $v_{\rm kick}=(20.3\pm1.5)~{\rm km/s}$. For model P2
the result is $E_{\rm rad}/M_{\rm ADM}=(0.0555\pm 0.0023)~\%$ and
$v_{\rm kick}=(19.7\pm 1.1)~{\rm km/s}$.


%
%

\section{Momentum flow}
\label{sec:Results}
In this section, we 
turn
to the momentum flow during the evolutions described in 
Sec.~\ref{sec:Simulations}. First, in Sec.~\ref{sec:MomentumSpEC} 
we measure 
the momentum of the holes during plunge, merger, and ringdown during 
a pseudospectral evolution of initial 
data set S1~(Table~\ref{Table:SpECID}), 
focusing on the momentum density and the inferred Landau-Lifshitz velocity 
$v_{\rm LL}^y$ along and opposite the frame-dragging direction 
(which in this paper are chosen to be the $\mp y$ direction, respectively). In
Sec.~\ref{sec:MomentumPuncture}, we look at the momentum flow in a 
moving-puncture simulation with similar initial data, 
and by comparing the puncture and spectral simulations, we 
investigate the influence of the 
choice of gauge on our results.
Then, in Sec.~\ref{sec:PN} we compare the momentum density and 
velocity of the holes with post-Newtonian predictions. 
\subsection{Pseudospectral results}\label{sec:MomentumSpEC}
\begin{figure}
\includegraphics[width=3.5in]{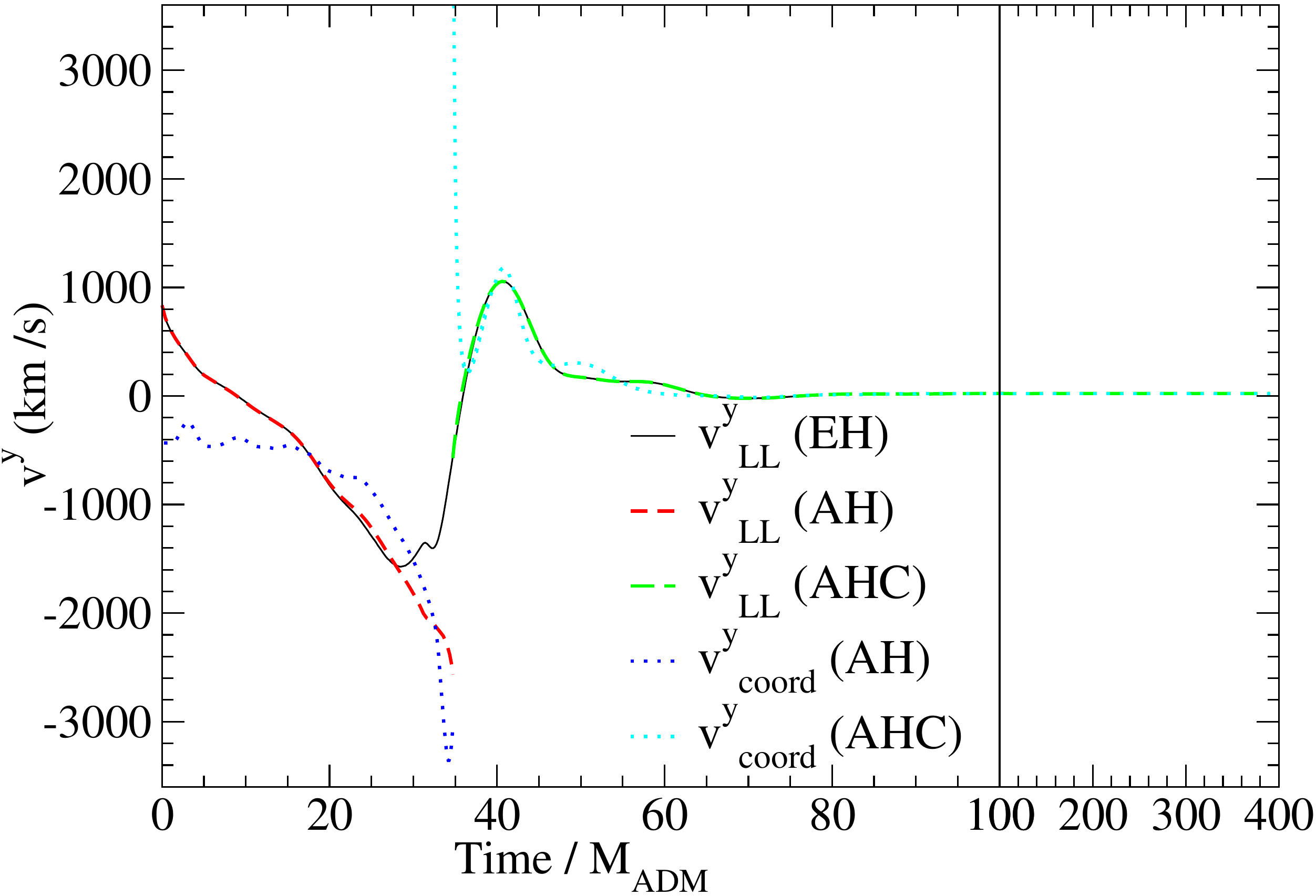}
\caption{
The velocity of the individual and merged black holes.
The Landau-Lifshitz velocity $v_{\rm LL}^y:=p^y_{\rm LL}/p^t_{\rm LL}$, 
where $p^\mu_{\rm LL}$ is the 
Landau-Lifshitz 4-momentum enclosed, is measured on the individual and common 
apparent horizons (labeled AH and AHC, respectively) 
and also on the event horizon (labeled EH). For comparison, 
the coordinate velocities $v^y_{\rm coord}$ of the apparent 
horizons are also shown. The data shown are from the high-resolution 
evolution $\mbox{N}2.C$.
\label{fig:LLAhOverview}\label{fig:SpECVel}}
\end{figure}

\begin{figure}[t!]
\includegraphics[width=3.5in]{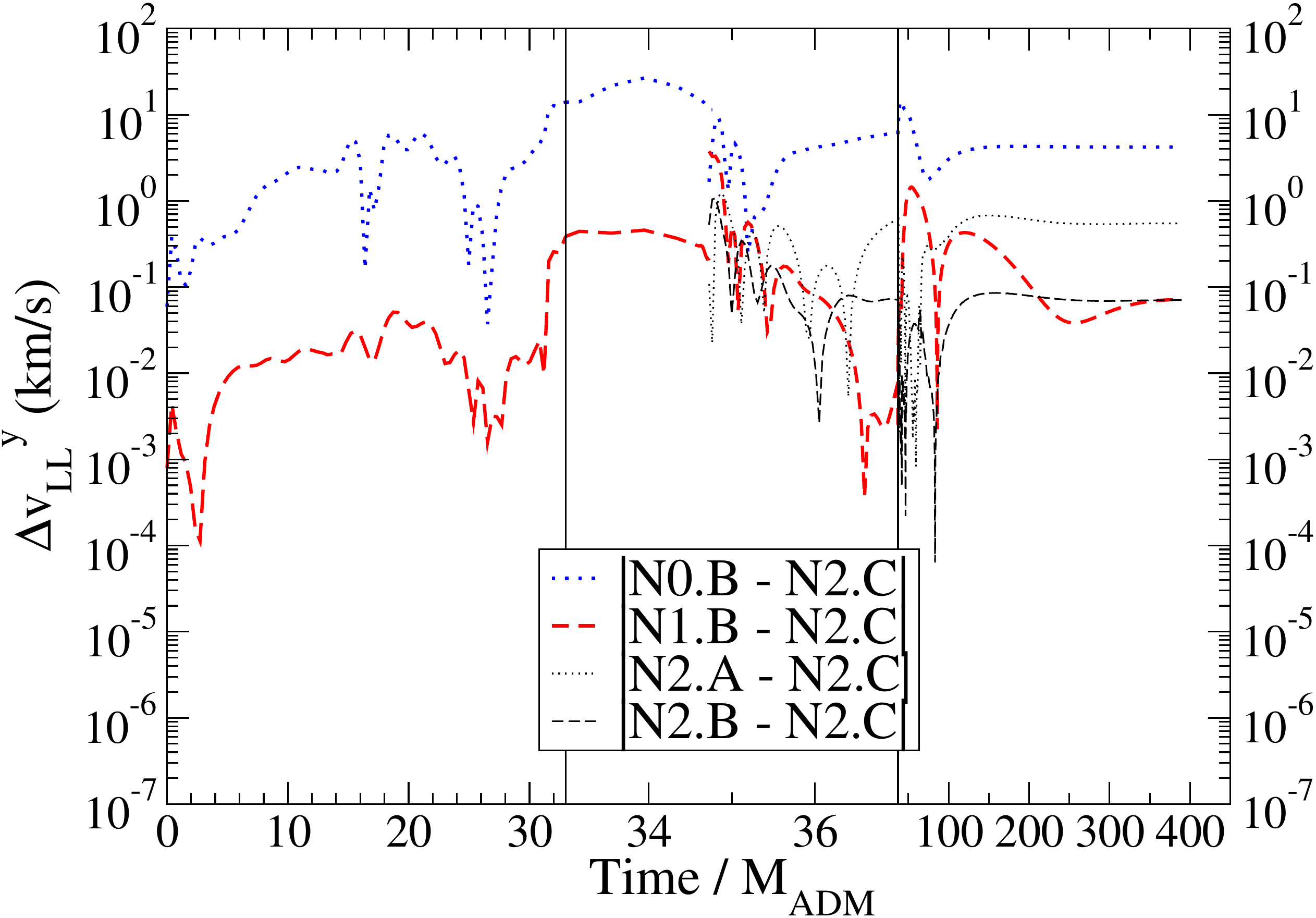}
\caption{\label{fig:Ev_3FC_vLLDiffVsHorizonFinderL} Convergence of  
$v_{\rm LL}^y$ with resolution. Specifically, differences between 
$v_{\rm LL}^y$ 
at the highest resolution $\mbox{N}2.C$ 
and at various lower resolutions are shown. 
Labels of the form $\mbox{N}x.y$ indicate the grid resolution, where
the pre-merger resolution is labeled (from coarse to fine) by $x=0,1,2$ 
and the post-merger resolution is labeled by $y=A,B,C$.
The 
difference between the second-highest and highest resolution is below 
$0.1 \mbox{km/s}$ except near merger, when it grows as large as 1 km/s.
\label{fig:vLLDiffVsResolution}}
\end{figure}

Throughout the pseudospectral evolutions summarized in 
Sec.~\ref{sec:EvolveSpEC}, we measure the 4-momentum density 
by explicitly computing the Landau-Lifshitz pseudotensor
[Eq.~(\ref{eq:LLPseudo})]. 
Because our evolution variables are essentially the spacetime 
metric $g_{\mu\nu}$ and its first derivative $g_{\mu\nu,\rho}$, we are 
able to compute the momentum density without taking any additional numerical 
derivatives. 
Besides measuring the 
momentum density, we also measure the 4-momentum 
$p^{\mu}_{A} $ [Eq.~(\ref{eq:pAsurf})] enclosed by i) the apparent horizons, 
ii) the event horizon, and iii) several spheres of large radius. 
From the enclosed momentum, we evaluate the effective velocity 
$v^j_{\rm LL}$ [Eq.~(\ref{eq:veff})]. 

\subsubsection{Apparent horizons}\label{sec:AHresults}

The effective velocities of the apparent horizons are shown in 
Fig.~\ref{fig:SpECVel} (dashed curves). 
To demonstrate convergence, 
Fig.~\ref{fig:vLLDiffVsResolution} shows the differences between 
apparent-horizon effective velocities computed at different resolutions.
During the plunge, the difference between the medium and fine resolution 
is less than 0.1 km/s until shortly before merger, when it reaches a few 
tenths of a km/s. Shortly after merger, the difference
between the highest 
and medium continuation resolutions between N2.B and N2.C
falls from about 1 km/s to about 0.1 
km/s.

For comparison, 
Fig.~\ref{fig:SpECVel} also shows the apparent horizons' coordinate 
velocities (dotted curves); the coordinate and effective velocities agree 
qualitatively during the plunge and quantitatively during the merger.
Also, Fig.~\ref{fig:SpECVel} shows that the effective velocities of 
individual apparent horizons and the the event horizon 
agree well until 
shortly before merger, when the event horizon's velocity smoothly 
transitions to agree with the common apparent horizon's 
(cf. Sec.~\ref{sec:EHresults} below). 

Because of frame-dragging, during the plunge 
the individual apparent horizons accelerate in the downward 
($-y$) direction, eventually reaching velocities of thousands of km/s.
But when the common apparent horizon appears, its velocity is much 
closer to zero and quickly changes sign, eventually reaching speeds of 
about 1000 km/s in the $+y$ direction (i.e., in the direction 
{\emph {opposite}} the frame-dragging 
direction). Then, as the common horizon rings down, the velocity relaxes 
to a final kick velocity of about 20 km/s in the $+y$ direction. 

\begin{figure*}
\begin{tabular*}{7in}{p{2.25in}p{2.25in}p{2.25in}}
\includegraphics[width=2in]{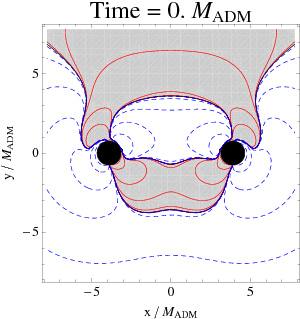}
& \includegraphics[width=2in]{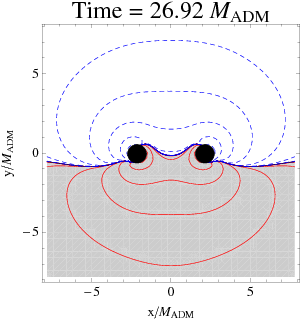}
& \includegraphics[width=2in]{Contour_LLty_Ev3FC_Lev4_F_Time89.png}  
\rule{0in}{2.25in}
\\
\includegraphics[width=2in]{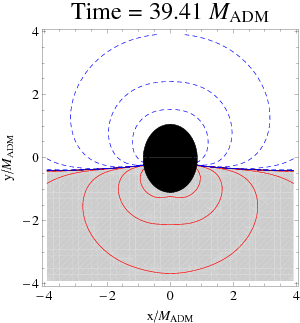}
& \includegraphics[width=2in]{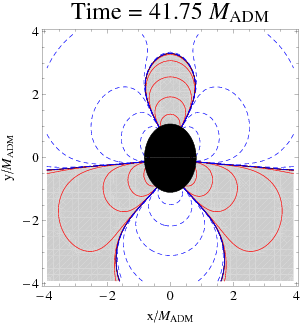}  
& \includegraphics[width=2in]{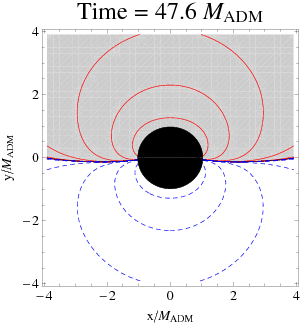}
   \rule{0in}{2.25in}
\\

& \includegraphics[width=2in]{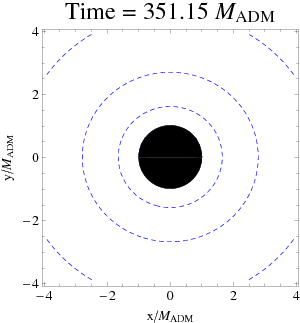}  
\rule{0in}{2.25in}
& 

\end{tabular*}
\caption{Contour plots of the $y$ (up-down) component 
of the momentum density, which points along or opposite of the holes' 
motion due to frame dragging. Adjacent contours correspond to 
a factor of 10 difference in the magnitude of the momentum density. 
Contours of positive $y$ momentum density are shown as solid red lines, 
while contours of negative $y$ momentum density are shown as 
dashed blue lines. The region containing positive $y$ momentum density 
is shaded grey. The regions inside the apparent horizons are 
shaded black, except for the upper right panel, where the region inside 
the \emph{individual} horizons is shaded black, while the common 
apparent horizon is indicated by a thick black line.
The data shown are from the high-resolution 
evolution $\mbox{N}2.C$.\label{fig:Contours}}
\end{figure*}

After merger, why have the horizon velocities suddenly changed from 
thousands of km/s in the frame-dragging direction to over a thousand km/s 
in the opposite direction? The answer can be seen in 
Fig.~\ref{fig:Contours}, which 
plots contours of constant y-momentum density at several times. 
At $t=0$, the momentum density has an irregular shape, because the 
initial data is initially not in equilibrium. By time $t=26.92 M_{\rm ADM}$, 
the momentum density 
has
relaxed. When the common apparent horizon forms 
(at time $t=34.73 M_{\rm ADM}$), 
it encloses not only the momentum of the 
individual apparent horizons but also the momentum in the gravitational field 
between the holes. 

\begin{figure*}
\flushleft
\includegraphics[width=3.4in]{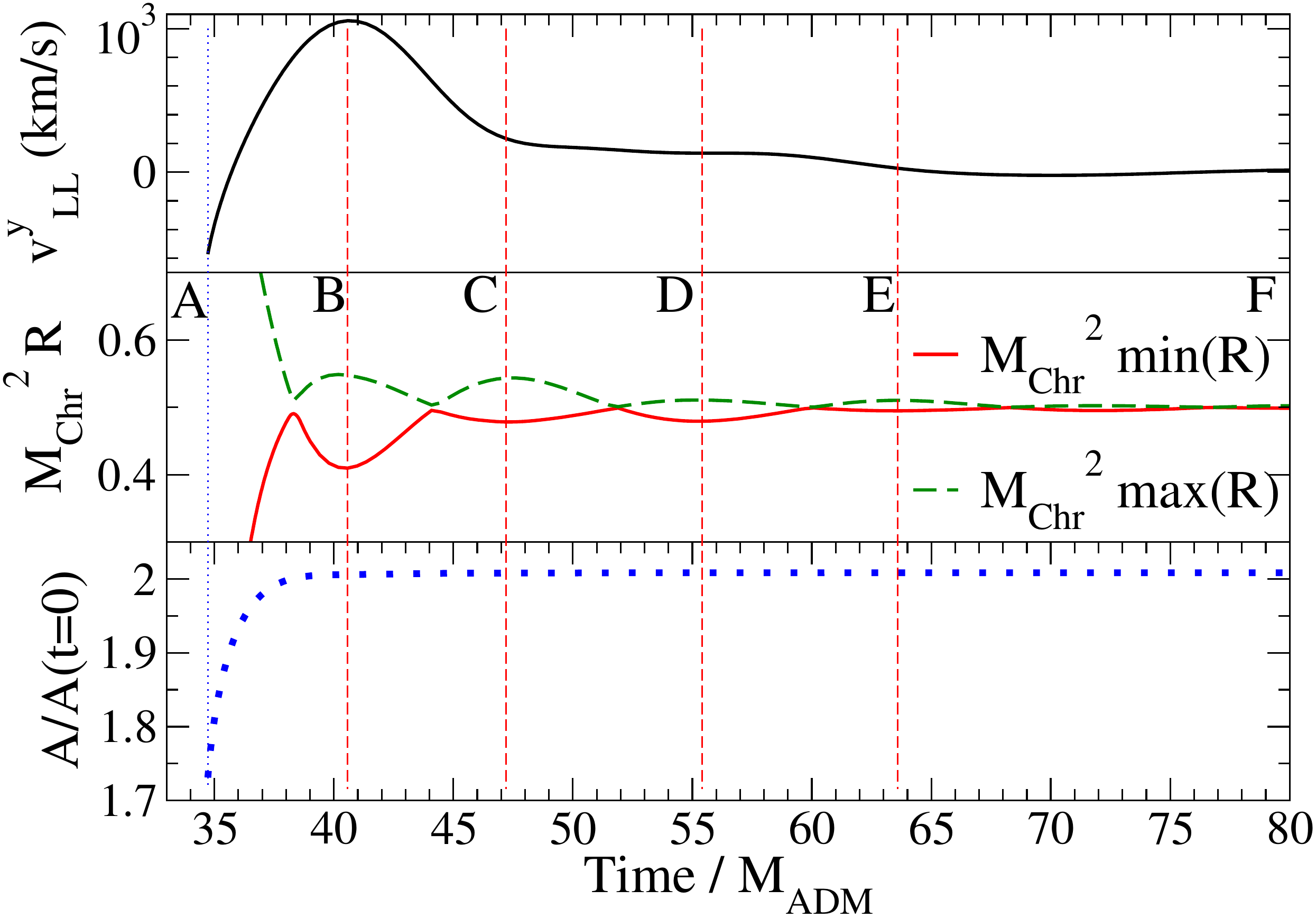} 
\includegraphics[width=3.4in, trim=0in 0.5in 0in 0in]{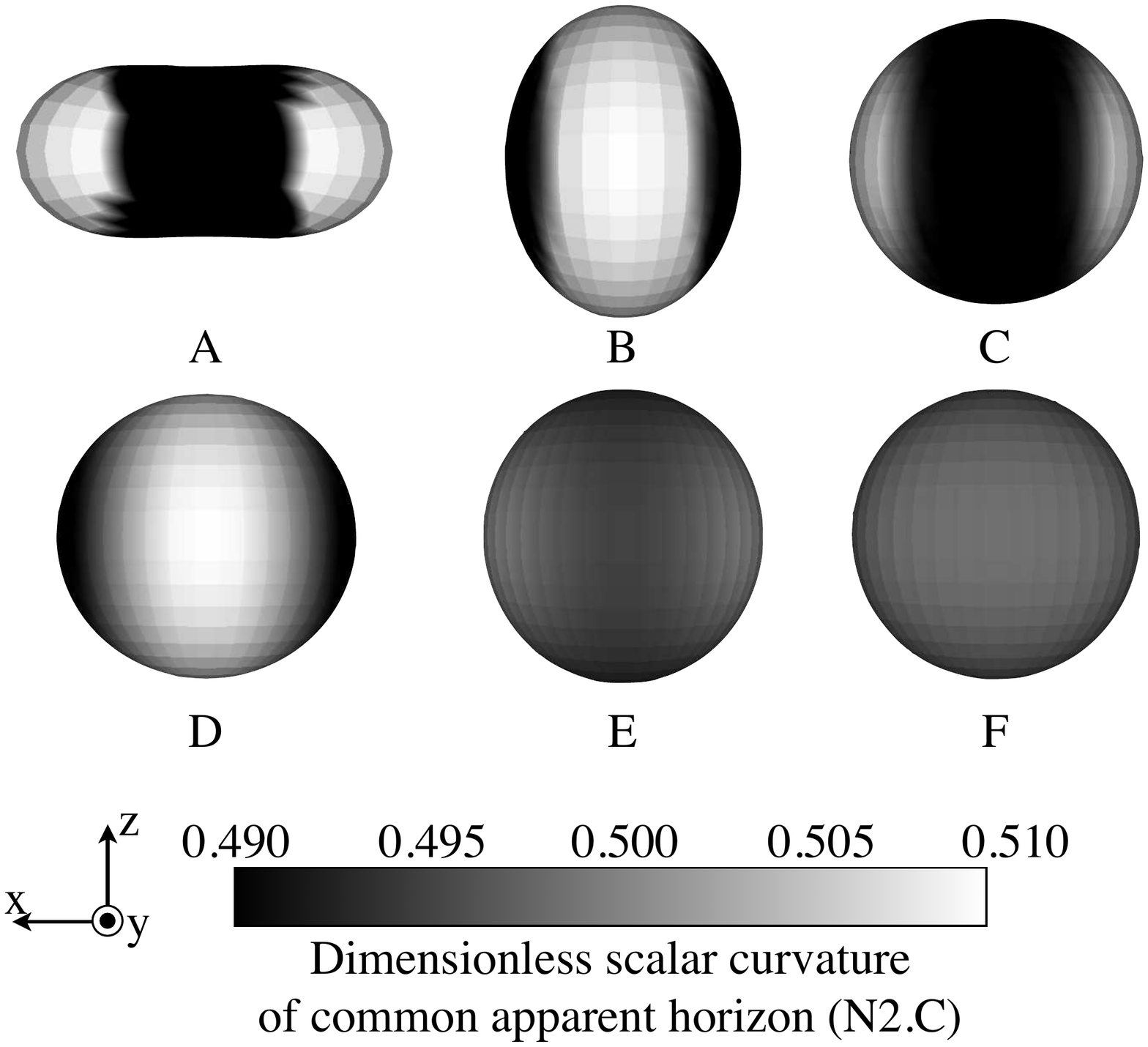} 
\caption{\emph{Left:} A comparison of the common apparent horizon's 
effective velocity and the horizon's shape and area. The top panel 
shows the horizon's effective velocity $v^y_{\rm LL}$. The middle panel 
shows the pointwise minimum and maximum of the horizon's 
dimensionless intrinsic scalar curvature; both $M_{\rm Chr}^2 \min(R)$ and 
$M_{\rm Chr}^2 \max(R)$ relax to the 
Schwarzschild value of $1/2$ as the horizon rings down. (The first four 
local minima of $M_{\rm Chr}^2 \min(R)$ are indicated by vertical dashed lines.) 
The bottom panel shows the area $A$ of the common apparent horizon 
normalized by the total area of the individual horizons at $t=0$. 
The data shown are from the high-resolution 
evolution $\mbox{N}2.C$. \emph{Right: } The dimensionless 
intrinsic scalar curvature $M_{\rm Chr}^2 R$ of the 
common apparent horizon at the times labeled A--F in the left panel. The 
horizon begins peanut-shaped, then rings down, eventually settling down to 
a sphere with a constant curvature $M_{\rm chr}^2 R=0.5$. 
\label{fig:Shape}}
\end{figure*}

It turns out that the net 
momentum outside the 
individual horizon but inside the common horizon points in the $+y$ 
direction; 
as the common horizon expands, it absorbs more and more of this upward 
momentum. 
Fig.~\ref{fig:Shape}
compares the common apparent horizon's effective velocity to its area and
shape; the latter is indicated 
by the pointwise maximum and minimum of the horizon's 
intrinsic scalar curvature. 
During the first half-period of oscillation (to the left of the leftmost dashed 
vertical line), the common horizon expands (as seen by its increasing area); 
as it expands, the
upward-pointing linear momentum it encloses causes 
$v_{\rm LL}^y$ to increase. After the first half-period, the 
horizon shape is maximally oblate 
(cf. panel B on the right side of of Fig.~\ref{fig:Shape}), 
and $v_{\rm LL}^y$ is at its maximum value of about $1000$ km/s. 

After another half-period of 
oscillation, the apparent horizon becomes prolate and encloses enough 
downward-pointing momentum that $v_{\rm LL}^{\rm eff}$ has decreased to only 
about $+200$ km/s.
After one additional full period, the effective 
velocity has fallen to nearly zero. As the horizon is ringing down, 
the momentum density in the surrounding gravitational field 
also oscillates: the final four panels 
in Fig.~\ref{fig:Contours} 
show how the momentum 
density relaxes to a final state as the horizon relaxes to that of a 
boosted Schwarzschild black hole. 

As the horizon rings down, 
gravitational waves are emitted, and these waves carry 
off a small amount of 
linear momentum. 
The net radiated momentum is only a small fraction of the momenta of the 
individual holes at the time of merger: 
the final effective velocity of the merged hole is 
about $20$ km/s in the upward-pointing direction, 
or about 1\% of the 
individual holes' downward velocity just before merger.

\begin{figure}[b!]
\includegraphics[width=3.5in]{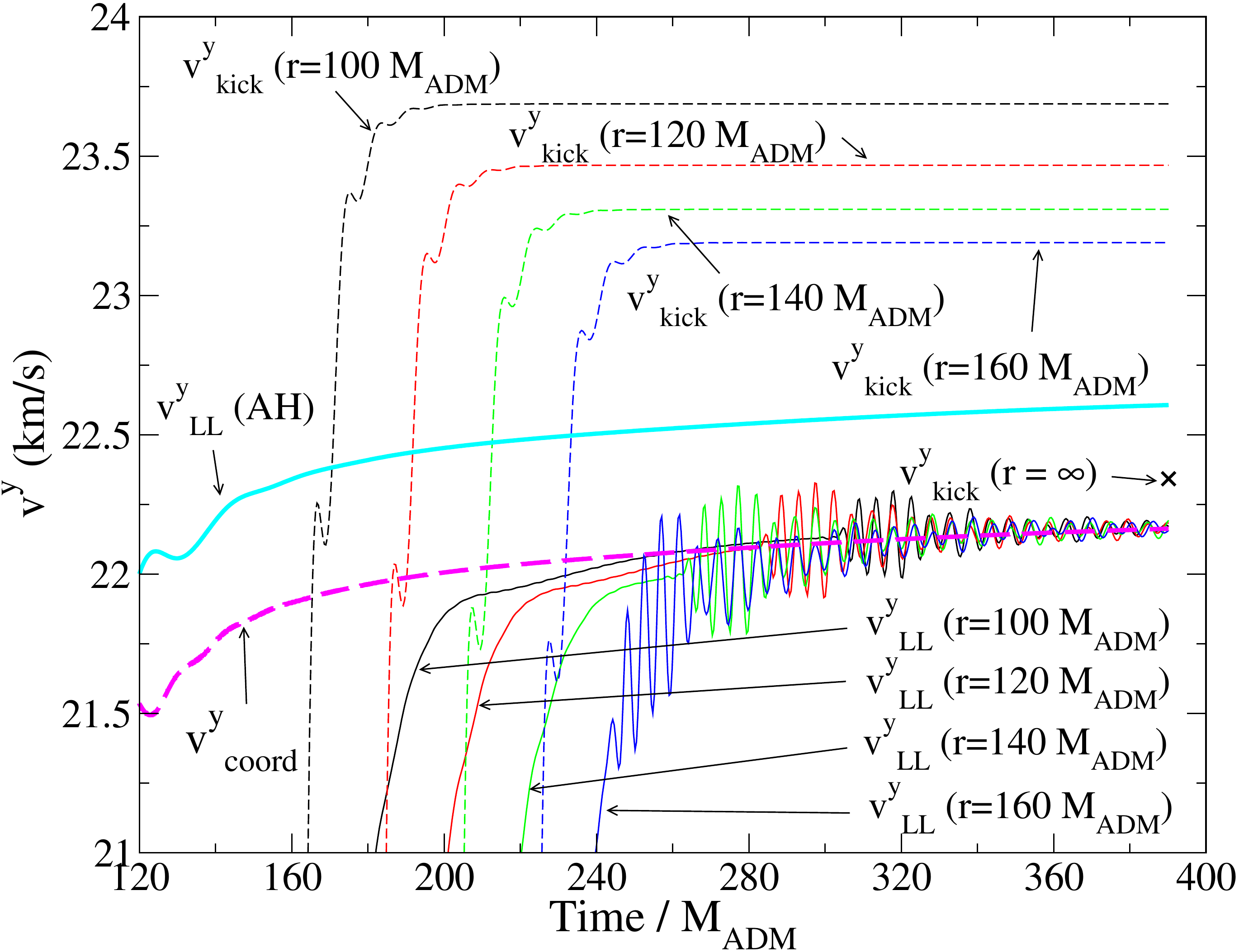}
\caption{A comparison of various measures of the merged hole's final velocity. 
The kick velocity inferred from the gravitational-wave flux 
(thin dashed lines) and the Landau-Lifshitz effective velocities 
$v^y_{\rm LL}$ (thin solid lines) are 
measured on spheres of radius $100 M_{\rm ADM}$, 
$120 M_{\rm ADM}$, $140 M_{\rm ADM}$, and $160 M_{\rm ADM}$
The value of the 
kick velocity at the final time is extrapolated to $r=\infty$ 
(black cross). The effective velocity measured on the common apparent horizon 
(thick solid line) and the coordinate velocity (thick dashed line) are also 
shown. 
The data shown are from the high-resolution 
evolution $\mbox{N}2.C$.\label{fig:SpECvelFinal}}
\end{figure}

Various measures of the final velocity of the merged hole are shown in 
Fig.~\ref{fig:SpECvelFinal}. The kick velocity $v^y_{\rm kick}$, 
which is inferred from the 
outgoing gravitational waves, is measured on 
four coordinate spheres 
(with radii $R$ of $100 M_{\rm ADM}$, $120 M_{\rm ADM}$, 
$140 M_{\rm ADM}$, and $160 M_{\rm ADM}$); the effective velocity 
is measured on the same coordinate spheres. We find that 
the effective velocity 
$v^y_{\rm LL}$ has no significant 
dependence on the radius of the extraction surface at late times, while
$v^y_{\rm kick}$ does. The dependence of $v^y_{\rm kick}$ on the
extraction radius is expected, since our method of extracting
$\Psi_4$ at finite radius has gauge-dependent contributions that vanish
as  $R\to\infty$.
When $v^y_{\rm kick}$ is extrapolated to infinite 
radius\footnote{To extrapolate, we fit the velocities 
$v^y_{\rm kick}$ at the final time to a function of radius $R$ of the form 
$a_0 + a_1/R + a_2/R^2$.}, 
however, it does agree well 
(within 0.2 km/s) with $v^y_{\rm LL}$. 
Also, the effective velocity $v^y_{\rm LL}$ calculated 
on the horizon also
agrees fairly well (within about 0.5 km/s) with $v^y_{\rm LL}$ measured 
on distant spheres. 

\subsubsection{Event horizon}\label{sec:EHresults}

\begin{figure}
\includegraphics[width=3.5in]{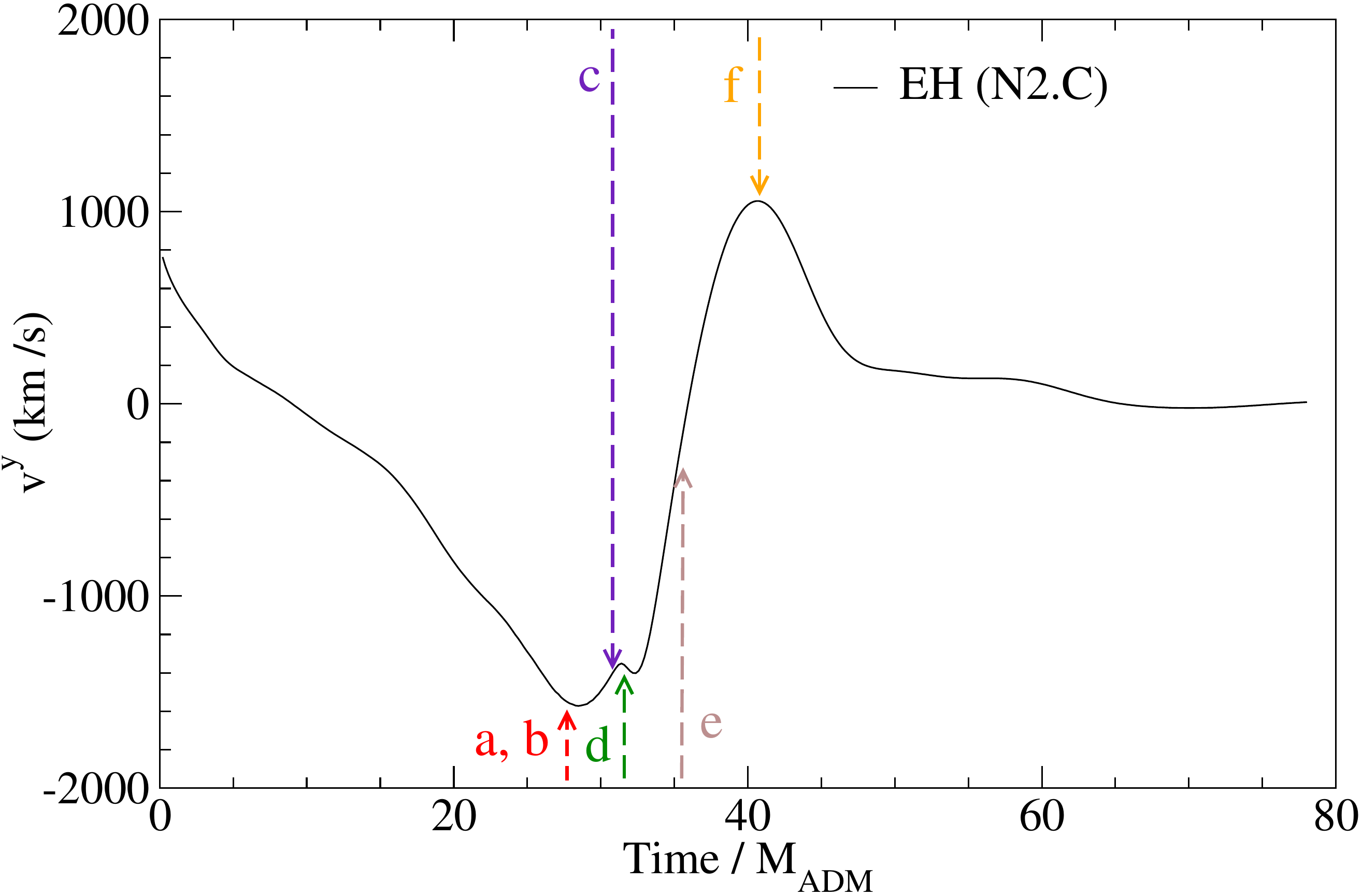}
\caption{ 
The effective velocity $v_{\rm LL}^y$ 
calculated on the event horizon surface, with the specified 
snapshots in Fig. \ref{fig:event_horizons} of the event horizon surface 
marked: a,b, $t = 27.7 M_{\rm ADM}$; c, $t = 30.8 M_{\rm ADM}$; d, 
$t = 31.6 M_{\rm ADM}$; e, $t = 35.5 M_{\rm ADM}$; f, $t = 40.8 M_{\rm ADM}.$
\label{fig:Ev_3FC_Vy_JustEH}}
\end{figure}

\begin{figure*}
\begin{tabular*}{12in}{p{3.00in}p{0.25in}p{3.00in}}
\includegraphics[width=2.4in]{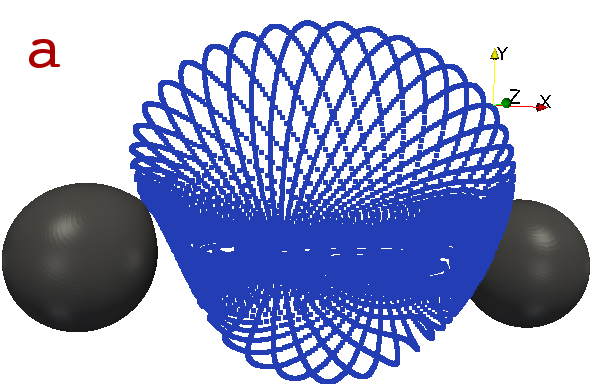}
&
& \includegraphics[width=2.4in]{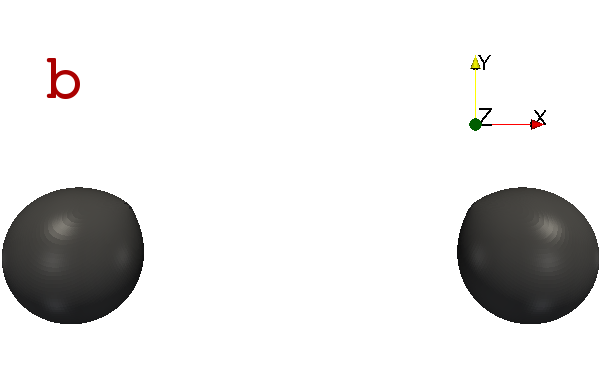}   \rule{0in}{2.00in}
\\

\includegraphics[width=2.4in]{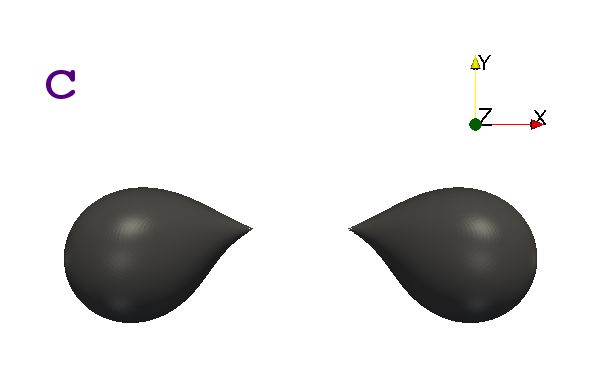}  
&
& \includegraphics[width=2.4in]{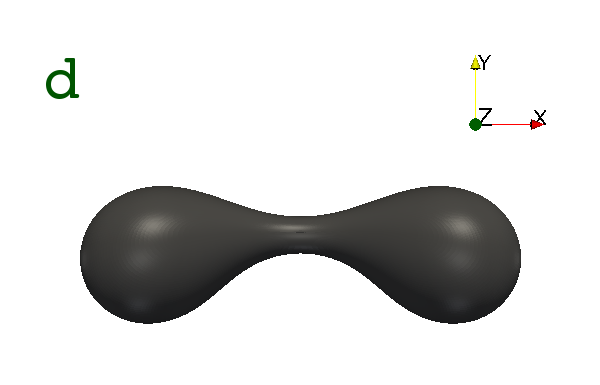}  \rule{0in}{2.00in}
\\
 \includegraphics[width=2.4in]{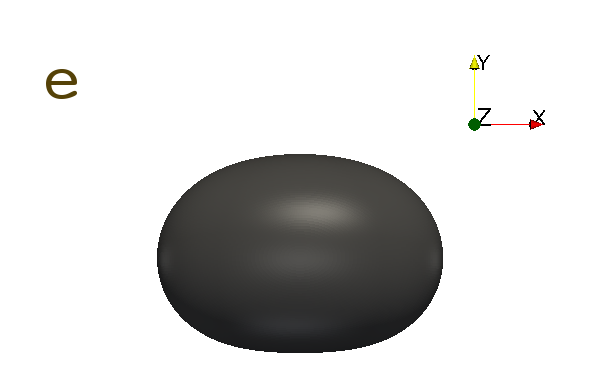}  
&
& \includegraphics[width=2.4in]{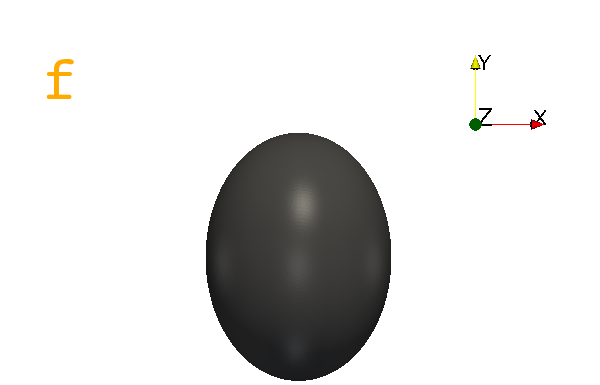}    \rule{0in}{2.00in}
 
\end{tabular*}
\caption{Snapshots of the event horizons at the times indicated in 
Fig.~\ref{fig:Ev_3FC_Vy_JustEH}: 
a,b, $t = 27.7 M_{\rm ADM}$; c, $t = 30.8 M_{\rm ADM}$; d, 
$t = 31.6 M_{\rm ADM}$; e, $t = 35.5 M_{\rm ADM}$; f, 
$t = 40.8 M_{\rm ADM}.$  All snapshots are looking down the
 z-axis to the x-y plane, except for shot a, which is slightly skewed
 (slightly rotated about the y axis) to better see the geodesic structure.
  In shot a, the future generators of the horizon are visible as small blue
 dots.  Note how the future generators
 map out a surface that meets the event horizon at
 the event horizon's cusps; this is where the future generators join the 
horizon. The data shown are from the high-resolution 
evolution $\mbox{N}2.C$.
\label{fig:event_horizons}}
\end{figure*}

We would like to compare our quantitative results of the 
effective velocity $v_{\rm LL}^y$ 
calculated 
using the event horizon surface (Fig.~\ref{fig:Ev_3FC_Vy_JustEH})
with qualitative observations of the event horizon's dynamics
(Fig.~\ref{fig:event_horizons}).  
We find that the greatest variation in \textit{both} the event horizon geometry
\textit{and} the value of  $v_{\rm LL}^y$ occurs
over a period of about 
$\Delta t = 13 M_{\rm ADM}$ from $t = 28 M_{\rm ADM}$ to $t = 41 M_{\rm ADM}$.
At time $t = 27.7 M_{\rm ADM}$, 
the cusps of the event horizon just begin to become noticeable 
(Figs.~\ref{fig:event_horizons} a \& b).  One can see in 
Fig.~\ref{fig:Ev_3FC_Vy_JustEH} that this is the time at which 
$v_{\rm LL}^y$ changes from decreasing to increasing. Shortly 
after\footnote{Note that at $t=31.1 M_{\rm ADM}$, 
we (smoothly) 
modify our gauge 
condition [Eq.~(\ref{eq:gaugehfalloff}) and the surrounding 
discussion]. The separate event horizons coalesce at 
time $t=31.1 M_{\rm ADM}$ as well; this is a coincidence.}, 
at $t = 31.1 M_{\rm ADM}$, the two separate event horizons coalesce into a 
common event horizon, and the common event horizon rapidly expands to form a 
convex shape by $t = 35.5 M_{\rm ADM}$ (Figs.~\ref{fig:event_horizons} d \& e). 
At this time, we note that $v_{\rm LL}^y$ is rapidly increasing 
(Fig.~\ref{fig:Ev_3FC_Vy_JustEH}, arrow e); this
rapid increase corresponds to the quickly expanding 
event horizon surface.

We 
interpret this process as the merging black holes 
``swallowing'' the gravitational field momentum between the holes.  
The resulting change in $v_{\rm LL}^y$ can be divided into two 
distinct portions: i) one that results from the changing 
event horizon surface 
in space, i.e. the field momentum swallowed by the black holes 
[mathematically, the \textit{second} term, in Eq.~(\ref{eq:dpAdtsurf})]
 and ii) a second that
 results from the change of field momentum at the black holes' surface, 
i.e. the field momentum \textit{flowing} into the black holes 
[mathematically, the \textit{first} term, in Eq.~(\ref{eq:dpAdtsurf})].  
While this 
distinction is clearly coordinate dependent, it 
could, after further investigation, nevertheless 
provide an intriguing and intuitive picture of the near-zone dynamics of
merging black hole binaries.

\subsection{Moving-puncture results and gauge}
\label{sec:MomentumPuncture}\label{sec:Gauge}

As summarized in Sec.~\ref{sec:LLformal}, the 
Landau-Lifshitz formalism that we have applied 
to our numerical simulations 
is based on a 
mapping between the curved spacetime of the simulation and an 
auxiliary flat spacetime. In the asymptotically-flat region far 
from the holes, there is a preferred way to construct this mapping.
Consequently, when the surface of integration 
is a sphere approaching infinite radius, Eq.~(\ref{eq:pAsurf}) gives a 
{\it gauge-invariant} measure of the system's total 4-momentum 
(see, e.g., Sec.~20.3 of Ref.~\cite{MTW}). However, when the surface 
of integration is in the strong-field region of the spacetime (e.g., 
when the surface is a horizon), the 4-momentum enclosed is gauge dependent. 
The momentum density, being given by a pseudotensor, is always gauge dependent.

The gauge-dependence of the effective velocity can be investigated 
at late times---when the spacetime has relaxed to its final, 
stationary configuration---by comparing the velocity obtained on the horizon 
with 
gauge-invariant measures of the kick velocity 
(Fig.~\ref{fig:SpECvelFinal}). 
At the 
final time in our pseudospectral simulation, the effective velocities
of the apparent and event horizons agree within tenths of a km/s with 
the (extrapolated) kick velocity inferred from the gravitational-wave flux;
at late times, the horizon effective velocities also agree 
with the effective velocity measured on 
coordinate spheres of large radius.
At least at late times, 
then, the 
effective velocity $v^y_{\rm LL}$ is not significantly affected by our 
choice of gauge.

But how strong is
the influence of gauge 
on our results in the highly-dynamical portion of the 
evolution, when we have no gauge-invariant measure of momentum or 
velocity?
To investigate 
this, 
we have 
evolved initial data that are physically similar  
using two manifestly different gauge conditions: 
i) the generalized-harmonic condition used in our spectral evolutions, 
and ii) the ``1+log'' slicing and ``Gamma-driver'' shift conditions used in 
our moving-puncture evolutions.

\begin{figure}
\includegraphics[width=3.0in]{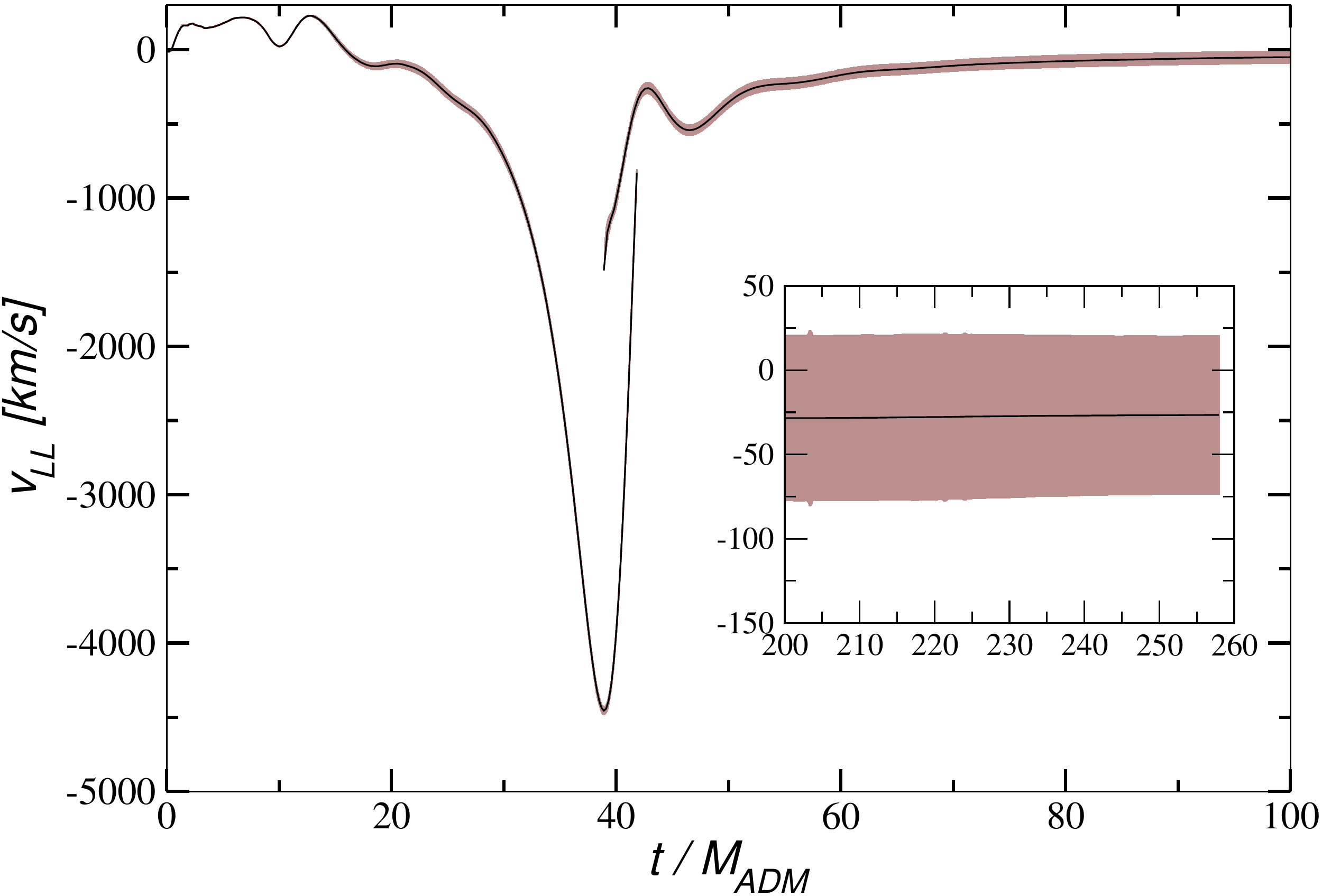}
\caption{Velocity perpendicular to the line of sight
         associated with the horizon integrals of the
         the Landau-Lifshitz tensor
         obtained for model P1. The shaded area represents
         the numerical uncertainty. During the pre-merger phase,
         the velocities of both holes are identical.
        }
\label{fig: punc_ll1}
\end{figure}

\begin{figure}
\includegraphics[width=3.0in]{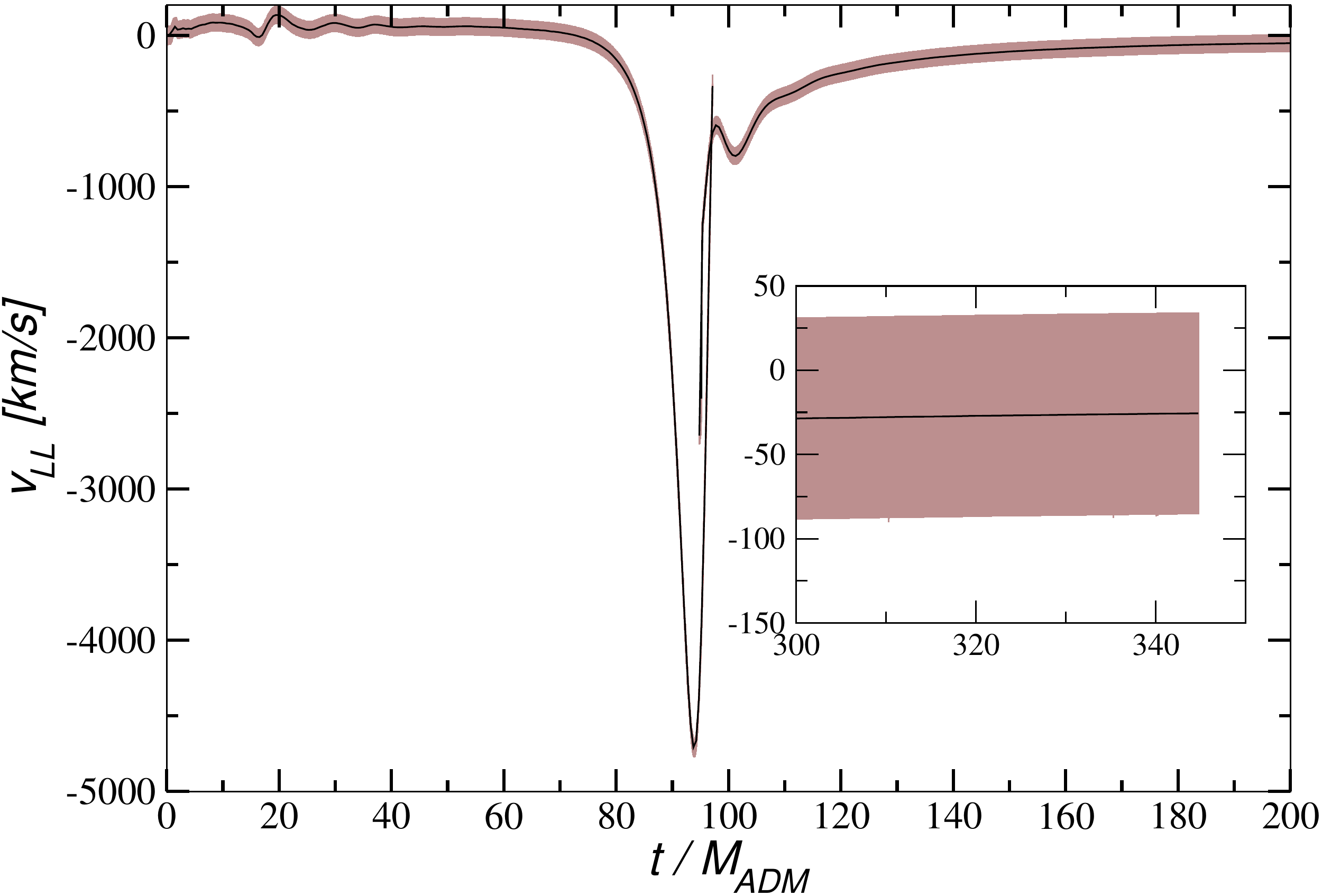}
\caption{Same as Fig.~\ref{fig: punc_ll1} for model P2 of
         Table \ref{tab: lean_models}.
        }
\label{fig: punc_ll2}
\end{figure}

Figs.~\ref{fig: punc_ll1} and \ref{fig: punc_ll2}
display the velocity obtained from the horizon integral of the
components of the Landau-Lifshitz tensor 
in the moving-puncture evolutions described in 
Sec.~\ref{sec:punctureEvolve}. 
The most remarkable
feature in these plots is a large temporary acceleration of the
black holes in the frame-dragging
direction. The magnitude of the velocity
reaches about $4500~{\rm km/s}$, which is of the order of the
superkicks first reported in Refs.~\cite{Gonzalez2007b, Campanelli2007a}.
In contrast to those inspiraling configurations, however, the black hole
motion reverses during the merger and settles down to a small value
of $-30 \pm 50~{\rm km/s}$. 


In order to examine to what extent this behavior is dependent on
specific properties of the puncture evolution 
(such as the particular form of the spurious radiation, which differs in 
our spectral and puncture evolutions), 
we have performed the
following additional simulations. First, we have changed the gauge
parameter $\eta$ in Eq.~(\ref{eq: shift}) to $0.75$ and $1.25$.
We do not observe a significant change in the behavior
of the effective velocity for this modification. 

Second, in order to gain further insight into the dependence of the
effective velocity on the initial separation of the black holes,
we have increased the initial separation of the holes
to allow for a longer pre-merger interaction phase;
We study the evolution of the second model P2 in
Table \ref{tab: lean_models}. This simulation has been performed with the
{\sc Lean} code as summarized in Sec.~\ref{sec: inidata_lean}
using a resolution $h_{\rm c}=M_{\rm ADM}/49.8$. The resulting
velocity is shown in Fig.~\ref{fig: punc_ll2} and
represents numerical uncertainties as gray shading.
The remarkable similarity between the figure and its
counterpart Fig.~\ref{fig: punc_ll1} for model P1 demonstrates that
the numerical results are essentially independent of the initial separation.

Comparing 
Figs.~\ref{fig:SpECVel} and \ref{fig: punc_ll1}, 
the qualitative behavior of the apparent horizons' 
effective velocities agrees. In both 
the spectral and puncture simulations:
\begin{enumerate} 
\item during the plunge, the individual apparent horizons accelerate 
to speeds larger than 1000 km/s in the frame dragging direction, 
\item when the common horizon forms, its velocity is much smaller in magnitude,
because 
the common horizon has enclosed momentum pointing opposite the frame-dragging 
direction, and 
\item the velocity relaxes to a value of only tens of km/s that 
(within numerical uncertainty) agrees with the kick velocity  
measured using the gravitational-wave flux.
\end{enumerate}

These results are particularly encouraging because two popular
gauge choices used in the NR community give remarkable overall
agreement. While this qualitative agreement certainly 
does not constitute a proof of
a gauge independence of our findings, we feel encouraged in our
hope that different types of observers might agree on their
overall perception of the local black-hole dynamics during the collision.
Most importantly from a practical point of view, it appears possible that
such local descriptions can be derived from the current generation
of BBH codes without the different numerical relativity groups
having to agree upon one and the same gauge choice for comparing their 
momentum densities and effective velocities. Future investigations
using a wider class of coordinate conditions should further clarify the
significance of gauge choices in this context.

\subsection{Comparison with post-Newtonian predictions}\label{sec:PN}

In this section we compare 
our results
to post-Newtonian predictions.
For each comparison, first the S1 data set (Table~\ref{Table:SpECID}) 
is presented
along with post-Newtonian predictions of a corresponding initial configuration,
then the H1 data set (Table~\ref{Table:SpECID}) is presented along with its
post-Newtonian predictions.  The post-Newtonian trajectories for spinning point
particles were
generated by evolving the post-Newtonian equations of
motion \cite{Faye-Blanchet-Buonanno:2006,Tagoshi-Ohashi-Owen:2001}.  The difference between the two data
sets are: i) set H1 begins with a larger initial separation than set S1, and
ii) set H1 is evolved in a nearly harmonic gauge.  
Comparing evolutions of data sets S1 and H1 
illustrates how these two effects improve the
comparisons one can make with post-Newtonian predictions.

\begin{figure*}
\includegraphics[width=3in]{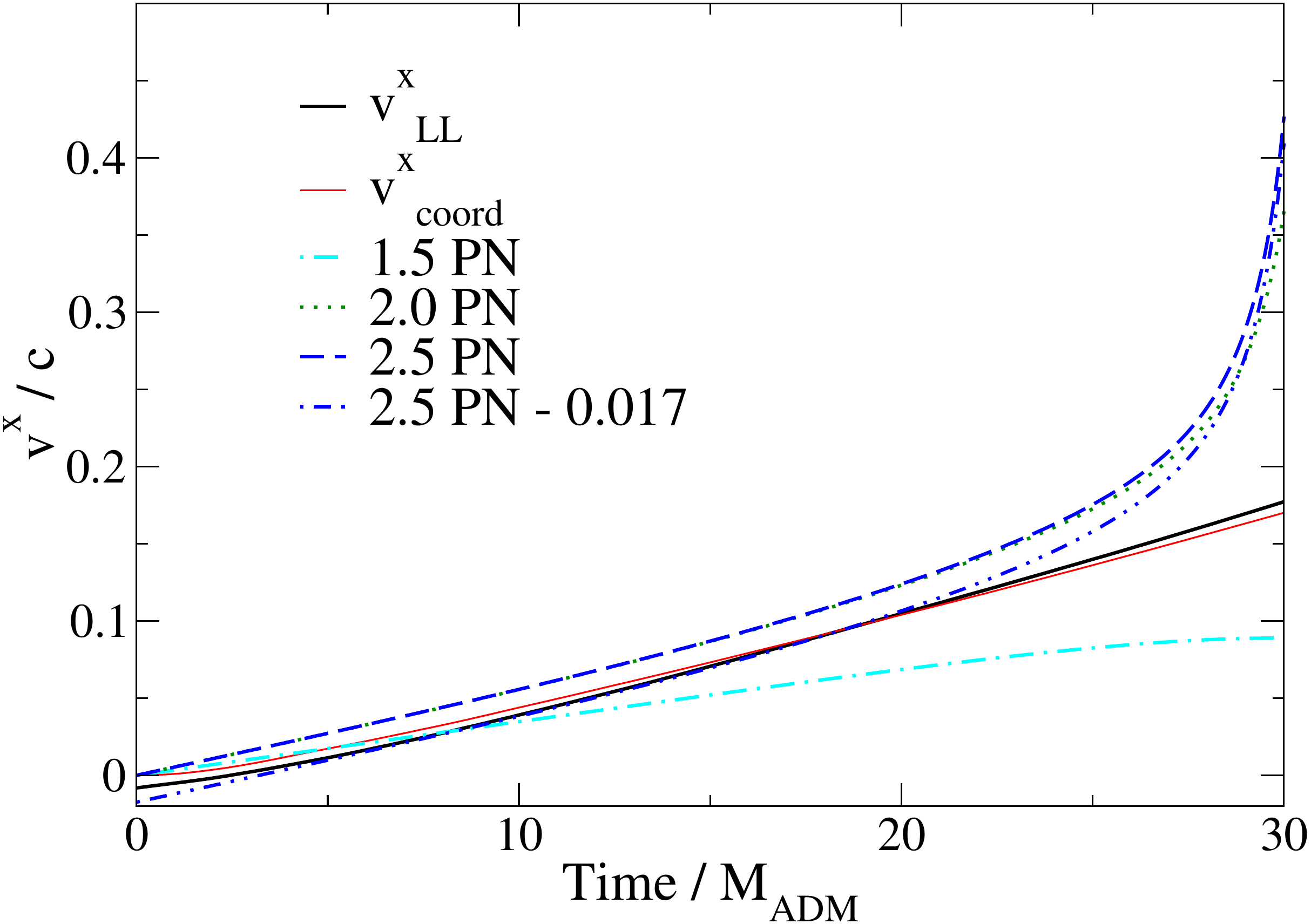}
\includegraphics[width=3in]{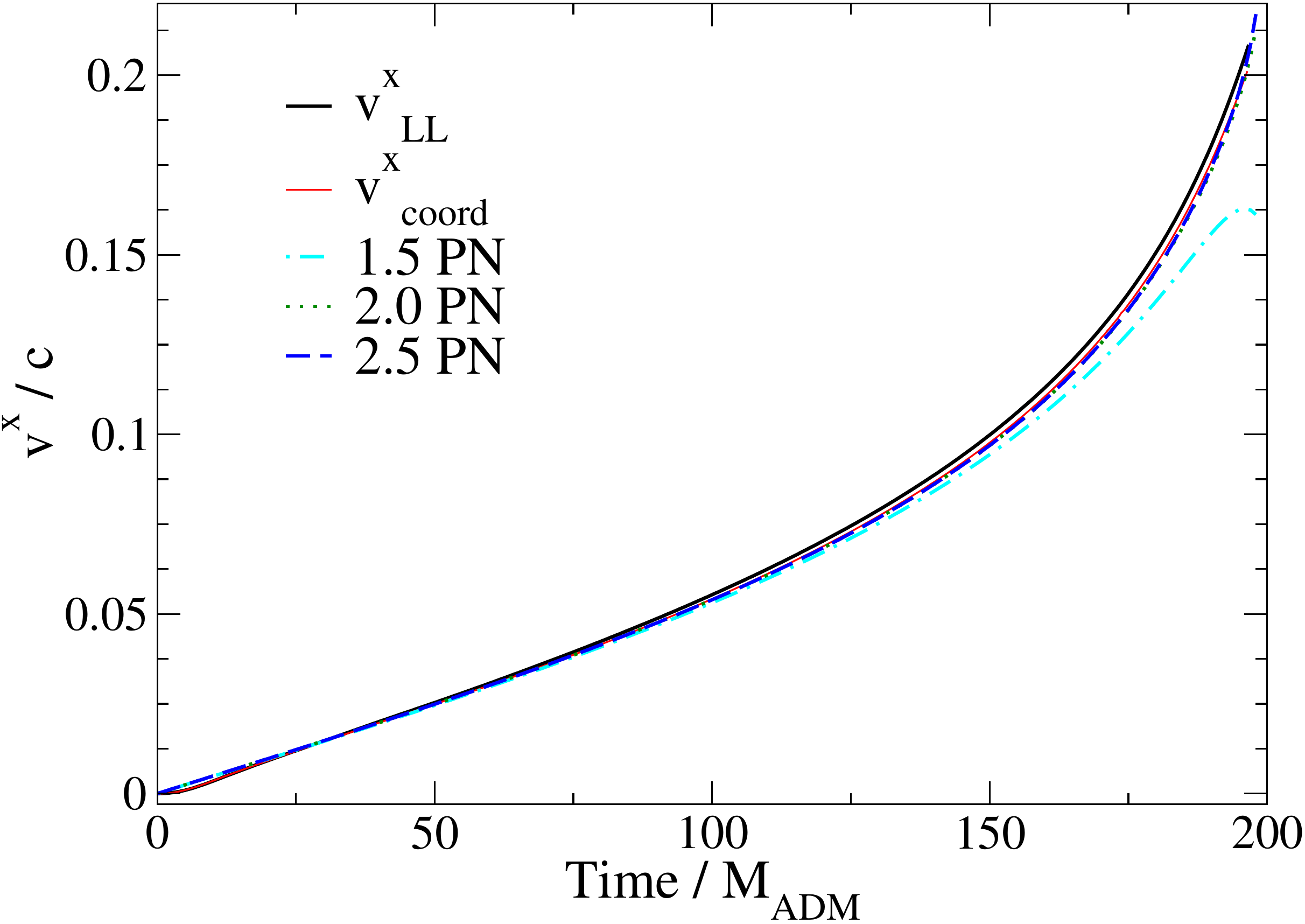}
\caption{
A comparison of numerical and post-Newtonian longitudinal velocities
(i.e., $v^x / c$) versus time.  The predicted coordinate velocities at several
post-Newtonian orders are shown as broken curves.
\emph{Left:} A comparison of S1 numerical data and post-Newtonian predictions. 
The numerical and post-Newtonian curves agree qualitatively. 
When the 2.5 PN curve is offset by a certain amount, 
it agrees quantitatively with the coordinate velocity $v^x_{\rm coord}$ and
the effective velocity $v^x_{\rm LL}$.
\emph{Right:} A comparison of H1 numerical data and PN predictions.
The effective velocity $v^x_{\rm LL}$ 
(thick black line) closely tracks the coordinate velocity $v^x_{\rm coord}$; 
both numerical curves also agree well with the 2.0 PN and 2.5 PN curves.
\label{fig:PNVxVsT}}
\end{figure*}

\begin{figure*}
\includegraphics[width=3in]{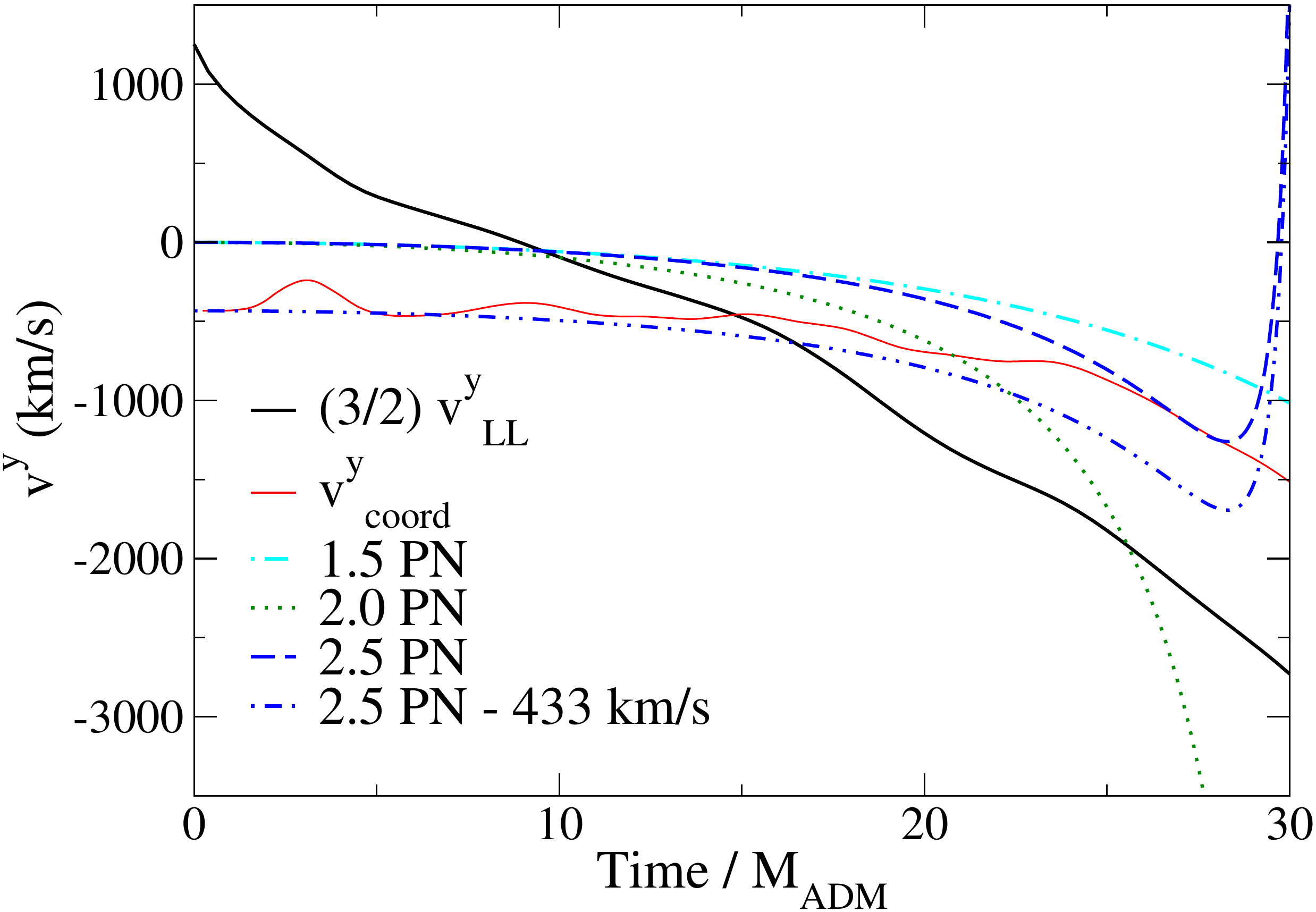}
\includegraphics[width=3in]{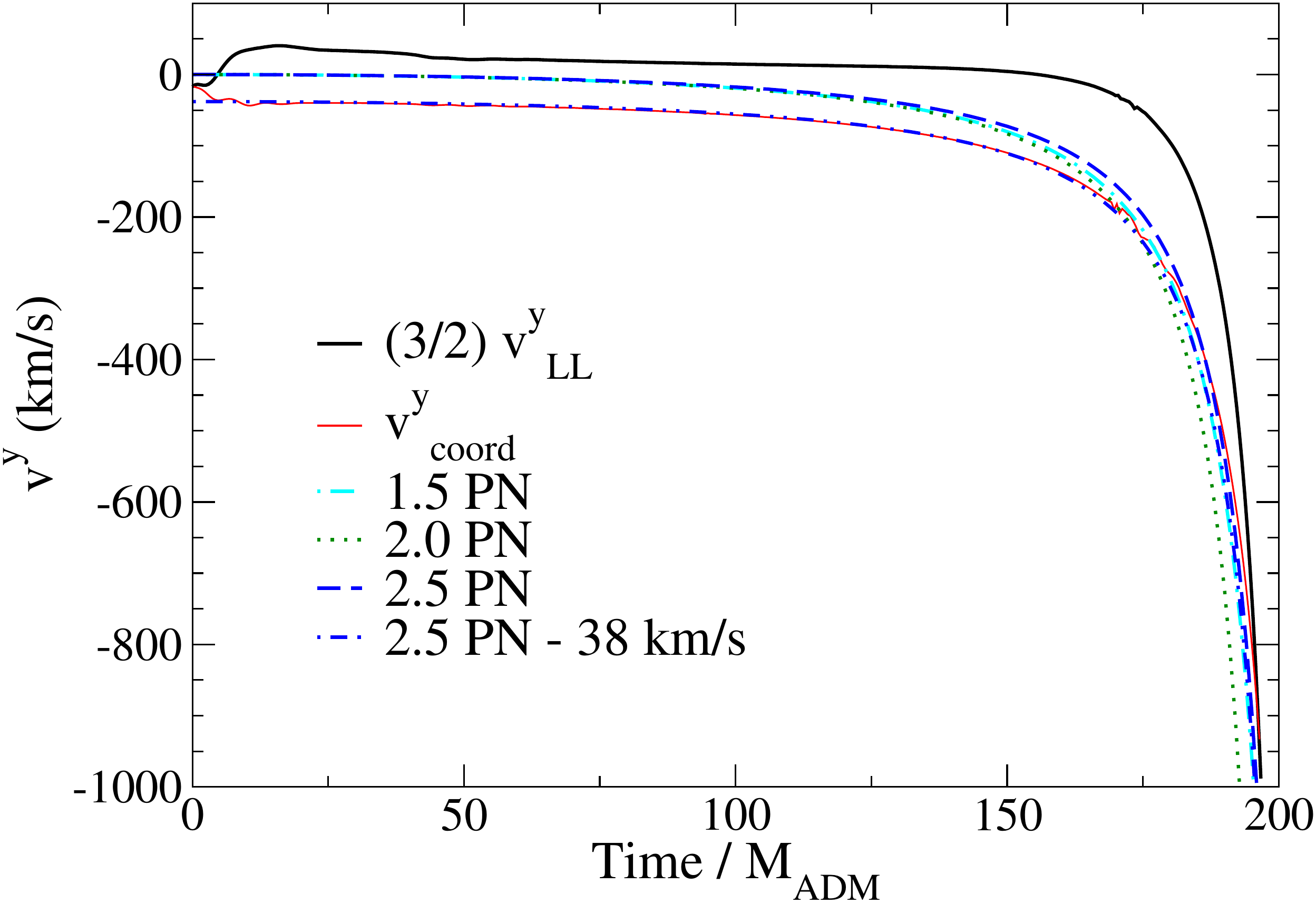}
\caption{
A comparison of numerical and post-Newtonian 
transverse velocities (i.e., $v^y$ in
km/s) versus time.  The left panel shows numerical results from 
simulation S1, 
while the right panel shows numerical results from simulation H1.
The predicted coordinate velocity at several
post-Newtonian orders are shown as broken curves.
The effective velocity is shown in black; it
has been rescaled by a factor of $3/2$ in order
to aid comparison with the post-Newtonian point-particle velocities,
as discussed in the text.
\label{fig:SHK_PNcmpr_VyVsT}
\label{fig:PNVyVsT}
}
\end{figure*}

\begin{figure*}
\includegraphics[width=3in]{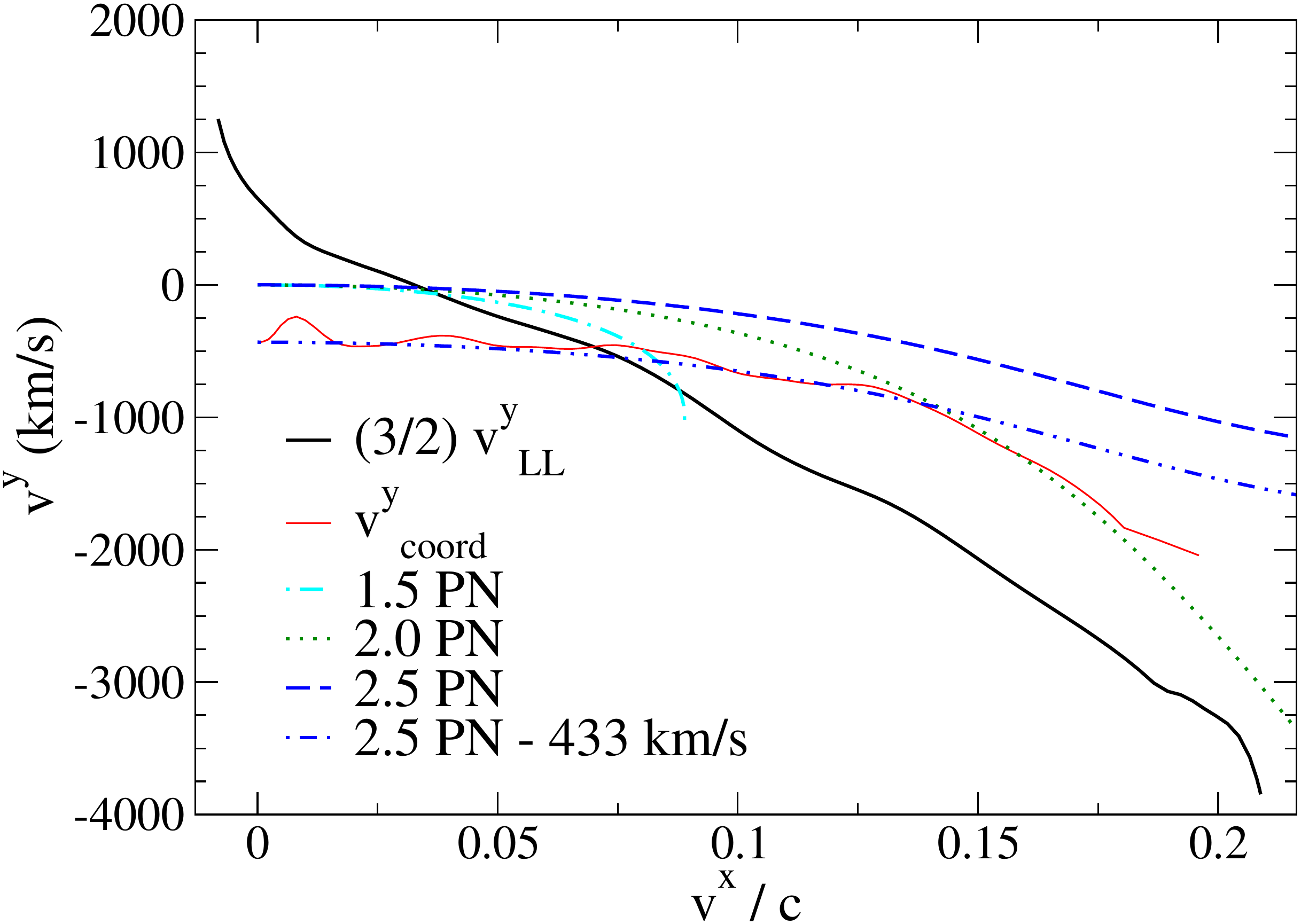}
\includegraphics[width=3in]{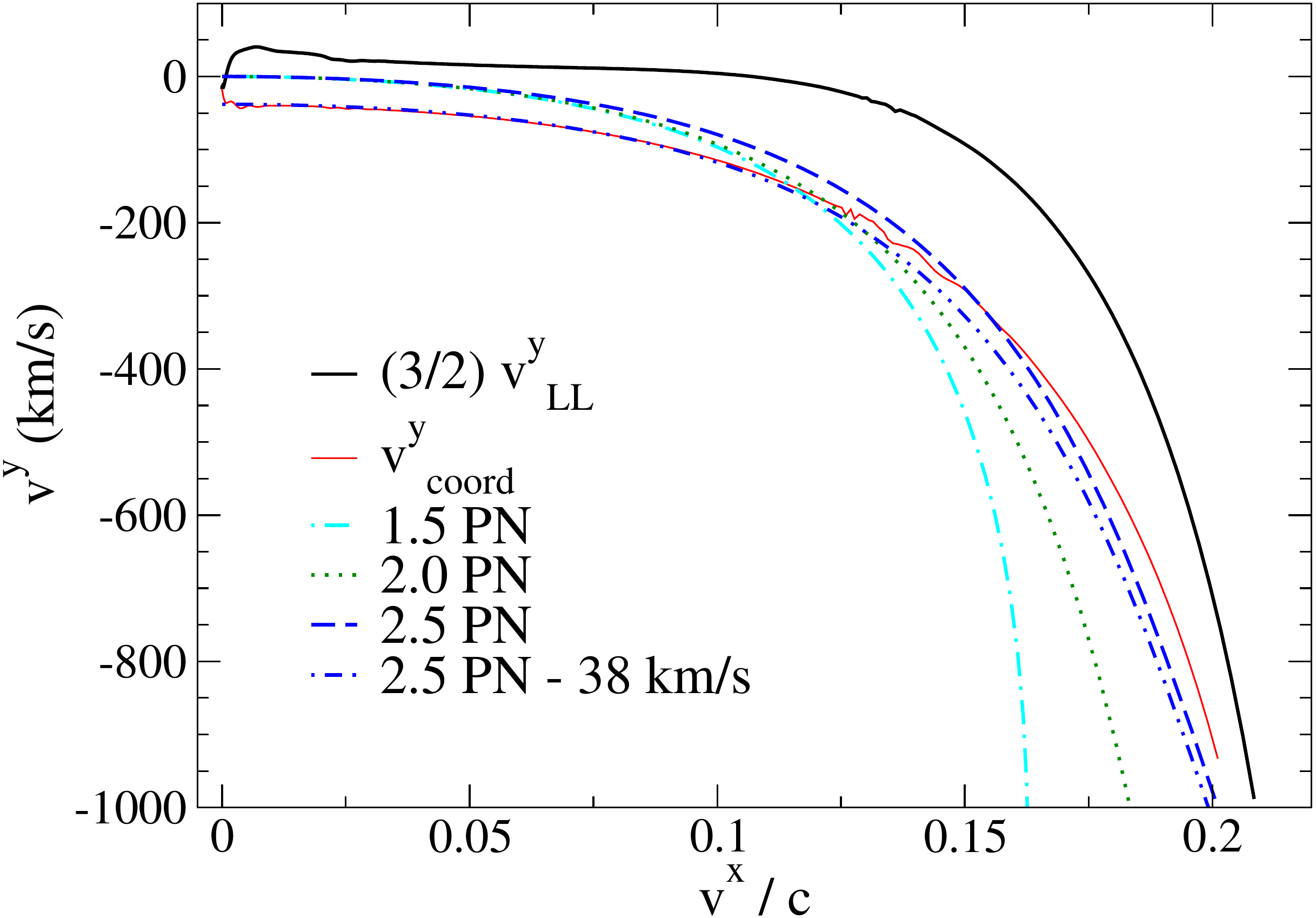}
\caption{
A comparison of numerical and post-Newtonian velocities.
In the figure, $v^y$ in km/s is plotted against $v^x/c$.
The effective velocity $v^y_{\rm LL}$ 
of the highest-resolution 
($\mbox{N}2.C$) 
evolution of initial data S1 (Table~\ref{Table:SpECID}) on the left and of the
evolution of initial data H1 (Table~\ref{Table:SpECID}) on the right are shown
as a 
thick black line. The predicted coordinate velocity at several post-Newtonian 
orders are shown as broken curves.
The transverse effective velocities only agree qualitatively with
post-Newtonian predictions; however, the coordinate velocity agrees very well
with post-Newtonian predictions.  In the left panel, the coordinate velocity 
has been artificially
truncated shortly before merger because at that point we do not have a 
good measure of the coordinate velocity.
The effective velocity has been rescaled by a factor of $3/2$ 
to aid comparison with the post-Newtonian point-particle velocities,
as discussed in the text.
\label{fig:PNVyVsVx}
\label{fig:SHK_PNcmpr_VyVsVx}
}
\end{figure*}

The
left panels of Figs.~\ref{fig:PNVxVsT}--\ref{fig:PNVyVsVx} 
show the comparison between the
highest-resolution evolution ($\mbox{N}2.C$) of initial data set S1 and several
orders of post-Newtonian predictions. The right panels of 
Figs.~\ref{fig:PNVxVsT}--\ref{fig:PNVyVsVx} show analogous comparisons 
with an evolution of initial data set H1. 

Figure \ref{fig:PNVxVsT} shows that the bulk, longitudinal motions (i.e.,
motion in the $x$ direction) agree both qualitatively and quantitatively with
post-Newtonian predictions through most of the plunge (i.e., a few $M_{\rm ADM}$
before the formation of the common apparent horizon) for both data sets. 
In the left panel of Fig.~\ref{fig:PNVxVsT}, we have added another 2.5 PN curve 
that is offset vertically such that the 2.5 PN coordinate
velocity agrees exactly
with the numerical effective velocity at $t \approx
18.34 M_{\rm ADM}$; this is done in order to account for the period of initial
relaxation in the S1 data set. Quantitative agreement is then found between 2.5
PN predictions and both the effective and coordinate velocities 
from $t\approx 5 M_{\rm ADM}$ through
$t \approx 20 M_{\rm ADM}$.  The right panel of Fig.~\ref{fig:PNVxVsT}, which
has less of an initial relaxation due to the increased separation, shows
excellent agreement between both the effective and coordinate velocities and
the 2.0 PN and 2.5 PN predictions.

For the minor (yet more interesting) transverse motion (i.e., the motion 
along the $y$ direction), we find only
qualitative agreement between the numerical data and post-Newtonian
predictions---spin-orbit coupling 
[more specifically, frame-dragging plus spin-curvature 
coupling, see Eq.~(5.11) of paper I and discussions thereafter] 
cause the holes to move in
the $-y$ direction during the plunge, 
reaching speeds of order 1000 km/s before the holes merge.  
The post-Newtonian expansion scheme we adopt 
(paper I and Refs.~\cite{Faye-Blanchet-Buonanno:2006,Tagoshi-Ohashi-Owen:2001}) 
uses a harmonic gauge, and a {\it physical} spin supplementary condition (SSC)
of $S^{\alpha\beta} u_\beta=0$, where $S^{\alpha\beta}$ is the 
spin angular momentum tensor of the black hole and $u^\beta$ its four velocity 
(see e.g., Sec.~II B of paper I).  

In this scheme, for the 
equal-mass--opposite-spin configuration, 
up to the leading 1.5\,PN order, the  coordinate $y$ velocity of the point 
particle representing each hole is equal to $3/2$ 
 the hole's effective velocity, $v^y_{\rm LL}$, evaluated through a 
surface integral of the post-Newtonian expression 
 for the super potential [cf.~Eq.~\eqref{eq:pAsurf}].  
Therefore, in Figs.~\ref{fig:PNVyVsT}--\ref{fig:PNVyVsVx} 
we rescale the effective velocity $p^y_{\rm LL}$ by 
this factor of 3/2, which arises 
from our particular
 choice of SSC and from field momentum distribution in the 
vicinity of the holes (see Secs.~II B and II C, and Table I of
 paper I for details).
 
In Figs.~\ref{fig:PNVyVsT} and \ref{fig:PNVyVsVx}, we 
compare the post-Newtonian point-particle $y$ velocity with 
the numerical coordinate $y$ velocity and $3/2$ of the 
numerical effective $y$ velocity $v^y_{\rm LL}$.
For the comparison to the S1
data set, we find qualitative agreement 
with both the effective and coordinate velocities
and the post-Newtonian predictions.  We think this agreement is not better 
because of
the large initial relaxations present in the S1 data set related to small
initial separation.  However, in the H1 
comparison, we find excellent
agreement between the coordinate velocity and the 2.5 PN prediction but only
qualitative agreement between the effective velocity and post-Newtonian
predictions. In these figures, 
offsets of $-$433\,km/s (for S1 data) and $-$38\,km/s (for H1 data) have 
been used to make 2.5\,PN coordinate velocity agree better with 
numerical results. 
Such offsets can be motivated as follows.
Our numerical initial data were chosen such that
the initial total momentum of the entire spacetime vanishes.  This, in our 
post-Newtonian scheme, corresponds to nonvanishing initial 
$y$ velocities of (see Table I of paper I)
\begin{equation}
v^y_{\rm coord} =\frac{\chi}{4 (r_0/M_{\rm ADM})^2}\,,
\end{equation}
where $\chi$ is the spin parameter of each  hole, and $r_0$ their 
initial separation. This corresponds to $-616$\,km/s
for the S1 data, and $-42$\,km/s for H1 data. 
  Again, the agreement is qualitative for
S1 data, and quantitative for H1 data.

\begin{figure*}
\begin{tabular*}{7in}{p{2.25in}p{2.25in}p{2.25in}}
\includegraphics[width=2in]{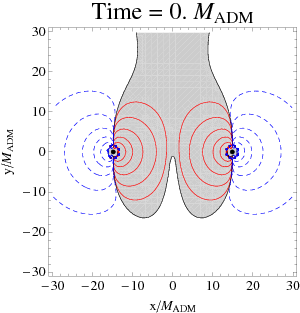}
& \includegraphics[width=2in]{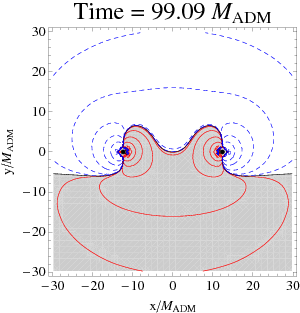}
& \includegraphics[width=2in]{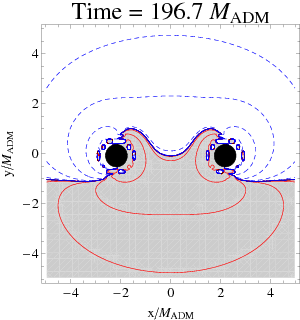}
\rule{0in}{2.25in}
\\
\includegraphics[width=2in]{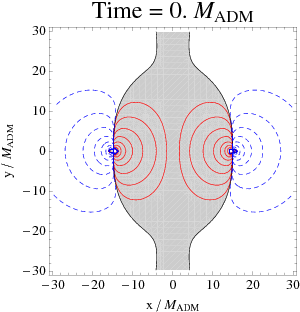}
& \includegraphics[width=2in]{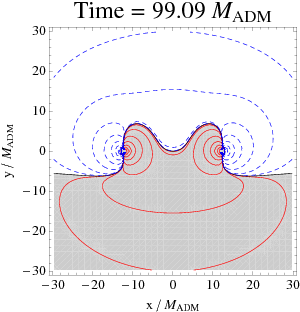}
& \includegraphics[width=2in]{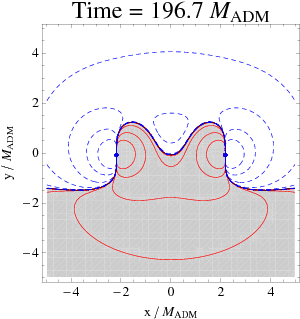}
\rule{0in}{2.25in}
\end{tabular*}
\caption{Comparison of numerical (top row) and post-Newtonian
    (bottom row) $y$ momentum density.  The numerical data comes from
    the harmonic evolution H1 described in
    Appendix \ref{sec:AppSHK}.
    The 1.5 PN momentum density is computed from Eqs.~(A2a)--(A2c)
    in paper I using the numerical hole trajectories.  As in
    Fig.~\ref{fig:Contours}, contours represent powers of $10$ in $y$
    momentum density.  The positive $y$ momentum density contours are
    shown in red, negative in blue.  The region of positive $y$
    momentum density is shaded grey.  In the numerical plots the
    apparent horizons are shown in black.}
\label{fig:H1_PN_Contours}
\end{figure*}

One final comparison we make between the H1 data set and post-Newtonian
predictions is the near-field momentum density, shown in
Fig.~\ref{fig:H1_PN_Contours}.  The numerical data comes from the harmonic
evolution H1, while the 1.5 PN momentum density is computed from
Eqs.~(A2a)-(A2c) in paper I using the numerical hole trajectories.  The left
panels, comparing the initial data to the predicted post-Newtonian momentum
density, show differences which are presumably due to 
differences in the post-Newtonian and numerical initial data, 
such as the numerical initial data being out of equilibrium.  
The center panels show the momentum densities agree very well once
enough time has elapsed for the 
spacetime to relax and for the spurious radiation to be emitted 
but
before the holes have fallen too close together.  The right panels make a final
comparison just before the holes get close enough to merge and shows
differences appearing between the numerical data and the post-Newtonian
predictions very near the holes---which could be an indication of the breakdown
of the post-Newtonian approximation.

These comparisons with post-Newtonian predictions 
have yielded several
interesting results.  The primary result 
of these comparisons is the surprisingly good agreement
found between post-Newtonian predictions and the coordinate velocities,
especially from the harmonic gauge evolution. Also, the longitudinal
effective and coordinate velocities track each other; consequently, the
longitudinal effective velocity agrees with post-Newtonian predictions.
The transverse effective velocities agree qualitatively with the post-Newtonian
predictions in the sense that they 
both indicate that the holes accelerate in the expected 
frame-dragging direction to speeds of order 1000 km/s. 
Finally, we have also found the
qualitative agreement between harmonic gauge numerical data and post-Newtonian
extends to the near-zone momentum density after the initial 
data relaxes but before the holes have fallen too close together.

\section{Conclusion}
\label{sec:Conclusions}

With the goal of 
building up greater physical intuition, we have used 
the Landau-Lifshitz momentum-flow formalism to 
explore the nonlinear dynamics of fully relativistic 
simulations of a head-on BBH plunge, merger, and ringdown. We have  
defined and computed an effective velocity of the black holes in terms of the 
momentum and mass-energy enclosed by their horizons, 
and we have interpreted the holes' transverse 
motion---which reaches speeds of order 1000 km/s---as a result of 
momentum flow between 
the holes 
and the gravitational field of the surrounding spacetime. We have 
found that the merged hole's 
final effective velocity---about 20 km/s---agrees with 
the recoil velocity implied by the momentum carried off by the 
emitted gravitational waves.

Our measures of linear momentum and effective velocity are 
gauge dependent. Nonetheless, after 
comparing simulations of comparable initial data in generalized-harmonic 
and moving-puncture gauges, 
we
have observed remarkably weak gauge dependence for the generalized-harmonic 
and moving-puncture evolutions discussed in this paper. 
Additionally, we have found surprisingly good agreement between the holes' 
effective and coordinate 
velocities, and at late times,
the holes' final effective velocities and gauge-invariant measures of the 
kick velocity agree.

These results motivate 
future explorations of momentum flow in 
fully-relativistic numerical simulations that are more 
astrophysically realistic. We are particularly eager to investigate 
simulations of 
superkick BBH mergers 
(the inspiral of a superkick configuration was considered using the 
post-Newtonian approximation 
in paper I).
Other future work includes 
studies of the linear and angular momentum flow in inspiraling 
(rather than head-on) 
mergers as well as mergers with larger spins.

\begin{acknowledgments}
We are pleased to acknowledge Michael Boyle, Jeandrew Brink, Lawrence Kidder,
Robert Owen, Harald Pfeiffer, Saul Teukolsky, 
and Kip Thorne for helpful discussions.
This work was supported in part by the Sherman Fairchild Foundation, 
the Brinson Foundation, the David and Barbara Groce Fund at Caltech, 
NSF grants PHY-0652952, DMS-0553677, PHY-0652929, 
PHY-0601459, PHY-0652995, PHY-0653653, DMS-0553302 and NASA grants 
NNX09AF96G and NNX09AF97G. 
Some calculations were done on the Ranger
cluster under NSF TeraGrid grant PHY-090003. 
\end{acknowledgments}

\appendix

\section{Superposed-Kerr-Schild (SKS) initial data}
\label{sec:AppID}
The initial data for the pseudospectral simulations presented in this paper 
was constructed using the methods described in Ref.~\cite{Lovelace2008}. 
In this appendix, we describe in more detail these initial data 
(which we summarize in 
Sec.~\ref{sec:SpECID}).

The usual 3+1 decomposition splits the spacetime metric $g_{\mu\nu}$ 
into a spatial metric $\gamma_{ij}$, lapse $\alpha$, and shift $\beta^i$, i.e.
\begin{eqnarray}
ds^2 & = & g_{\mu\nu}dx^\mu dx^\nu = -\alpha^2 dt^2 \nonumber\\
& + & \gamma_{ij}(dx^i+\beta^i dt)(dx^j+\beta^j dt).
\end{eqnarray} On the initial spatial slice (at time $t=0$), the initial data 
must specify the spatial metric $\gamma_{ij}$ and the extrinsic curvature 
$K_{ij}$, which is related to the time derivative of the spatial metric by
\begin{eqnarray}
\partial_t \gamma_{ij} = -2 \alpha K_{ij} + 2\nabla_{(i}\beta_{j)}.
\end{eqnarray} 

We use the quasiequilibrium 
formalism~\cite{Cook2002,Cook2004,Caudill-etal:2006,Gourgoulhon2001,
Grandclement2002}, in which $\gamma_{ij}$ and $K_{ij}$ are expanded as
\begin{eqnarray}
\gamma_{ij} = \psi^4 \tilde{\gamma}_{ij},\nonumber\\
K_{ij} = A_{ij} + \frac{1}{3}\gamma_{ij} K.
\end{eqnarray} The conformal metric $\tilde{\gamma}_{ij}$, the trace of the 
extrinsic curvature $K$, and their time derivatives can be chosen freely.
We adopt the quasiequilibrium choices
\begin{eqnarray}
\tilde{u}_{ij}:=\partial_t \tilde{\gamma}_{ij} = 0,\nonumber\\
\partial_t K = 0.
\end{eqnarray} The remaining free data are based on a weighted superposition 
of two boosted, spinning Kerr-Schild black holes (Eqs.~(45)--(46) of 
Ref.~\cite{Lovelace2008}):
\begin{eqnarray}\label{eq:CMetricSKS}
\tilde{\gamma}_{ij} & := & f_{ij}
+\sum_{a=1}^{2}e^{-r_a^2/w_a^2}\left(\gamma_{ij}^a - f_{ij}\right),\\
K & := & \sum_{a=1}^{2} e^{-r_a^2/w_a^2}K_a.\label{eq:TrKSKS}.
\end{eqnarray} Here $f_{ij}$ is the metric of flat space, $r_a$ is the 
Euclidean distance from the center of the apparent horizon of hole $a$, 
and $\gamma_{ij}^a$ and $K_a$ are the spatial metric and mean curvature of 
a boosted (with velocity $\tilde{v}^i$), spinning 
(with spin $\tilde{S}/\tilde{M}^2$) Kerr-Schild black hole 
centered at the initial 
position of hole $a$. In this paper we choose $\tilde{v}^i=0$ (since we 
seek data describing holes falling head-on from rest), 
$\tilde{M}=0.39 M_{\rm ADM}$, 
and 
$\tilde{S}/\tilde{M}^2=0.5$. The Gaussian weighting parameter  
is chosen to be $w_a=d/3$, where $d$ is the initial coordinate separation 
between the two holes; note that this choice causes the conformal metric 
to be flat everywhere except near each hole. 
The holes are located at coordinates 
$(x,y,z)=(x_0\equiv\pm d/2,0,0)$.

These free data are then inserted into the extended conformal thin sandwich 
(XCTS) equations (e.g., Eqs.~(13)--(15) of 
Ref.~\cite{Cook2004})\footnote{The XCTS equations are also
given by Eqs.~(37a)--(37d) of Ref.~\cite{Lovelace2008}, aside from the 
following typographical error:
the second term 
in square brackets on the right-hand-side of Eq.~(37c) 
should read $(5/12) K^2 \psi^4$ (not $(5/12) K^4 \psi^4$).
}, which are then solved 
for the conformal factor $\psi$, the lapse $\alpha$, 
and the shift $\beta^i$:
\begin{eqnarray}\label{junkeq:XCTSa}
\CCDu^2\CF - \frac{1}{8}\CRicciS\CF & - & \frac{1}{12}\TrExCurv^2\CF^5 
+ \frac{1}{8}\CF^{-7}\CA^{ij}\CA_{ij} = 0,\nonumber\\
\label{junkeq:XCTSb}
\CCD_j\Big(\frac{\psi^7}{2(\alpha\psi)}\CLong{\Shift}^{ij}\Big)
& - & \frac{2}{3}\CF^6\CCDu^i\TrExCurv
-\CCD_j\Big(\frac{\psi^7}{2(\alpha\psi)}\dtCMetric^{ij}\Big)= 0,\nonumber\\
\CCDu^2(\Lapse\CF) & - & (\Lapse\CF)\bigg[\frac{\CRicciS}{8}\!+\!\frac{5}
{12}\TrExCurv^2\CF^4\!
+\!\frac{7}{8}\CF^{-8}\CA^{ij}\CA_{ij}\bigg]
\label{junkeq:Lapse2}\nonumber\\
& = & -\CF^5(\dtime\TrExCurv-\Shift^k\partial_k\TrExCurv),
\label{junkeq:XCTSc}
\end{eqnarray} where 
the Ricci scalar curvature of the 
conformal metric $\tilde{\gamma}_{ij}$ is $\tilde{R}$, 
the longitudinal derivative 
$\tilde{\mathbb{L}}$ is defined as 
\begin{equation}
(\tilde{\mathbb{L}}V)_{ij} := \tilde{\nabla}_i V_j 
+ \tilde{\nabla}_j V_i - \frac{2}{3} \tilde{\gamma}_{ij} \tilde{\nabla}_k V^k,
\end{equation} and the trace-free part of the extrinsic curvature $A^{ij}$ 
satisfies 
\begin{equation}
\tilde{A}^{ij} = \psi^{10} A^{ij} 
= \frac{\psi^7}{2 (\alpha\psi)}
\left[(\tilde{\mathbb{L}}\beta)^{ij}-\tilde{u}^{ij}\right]\nonumber.
\end{equation} 

The XCTS equations are solved using 
a spectral elliptic solver~\cite{Pfeiffer2003}
on a computational domain 
with i) a very large outer boundary (which is chosen to be a coordinate 
sphere with radius $10^9 \tilde{M}$), and ii) with the region inside the holes' 
apparent horizons excised. The excision surfaces $\mathcal{S}$ are surfaces 
of constant Kerr radius $r_{\rm Kerr}$, where
\begin{eqnarray}
\frac{x^2+y^2}{r_{\rm Kerr}^2+\tilde{S_a}^2/\tilde{M_a}^2}
+\frac{z^2}{r_{\rm Kerr}^2}=1.
\end{eqnarray} The excision surfaces are the apparent horizons of the 
holes; this is enforced by the following boundary condition: 
(Eq.~(48) of Ref.~\cite{Cook2004}):
\begin{eqnarray}
\CSpatialNormal^k\partial_k \CF & = & 
-\frac{\CF^{-3}}{8\CLapse}\CSpatialNormal^i\CSpatialNormal^j
\left[\CLong{\Shift}_{ij}-\dtCMetric_{ij}\right]\nonumber\\
&&-\frac{\CF}{4}\,\CTwoMetric^{ij}\CCD_i\CSpatialNormal_j
+\frac{1}{6}\TrExCurv\CF^3 \mbox{ on }\mathcal{S}.
\end{eqnarray} Here $\tilde{s}^i:=\psi^2 s^i$, 
$s^i$ is a unit vector normal to the 
excision surface $\mathcal{S}$, and
$\tilde{h}_{ij}:=\tilde{\gamma}_{ij}-\tilde{s}_i\tilde{s}_j$ is the 
induced metric on the excision surface.

On the apparent horizon, the lapse satisfies the boundary 
condition
\begin{eqnarray}\label{eq:IDLapseBC}
\alpha\psi = 1 + \sum_{a=1}^2 e^{-r_a^2/w_a^2} (\alpha_a-1 )
\mbox{ on }\mathcal{S},
\end{eqnarray} where $\alpha_a$ is the lapse of the Kerr-Schild metric 
corresponding to hole $a$.
The shift satisfies
\begin{eqnarray}\label{eq:IDShiftBC}
\Shift^i = \Lapse s^i - \Omega_r \xi^i\mbox{ on }\mathcal{S}.
\end{eqnarray} The first term in Eq.~(\ref{eq:IDShiftBC}) implies that the 
holes are initially at rest, and the second term determines the spin of the 
hole; to make the spin point in the $\pm z$ direction with  
magnitude $S/M_{\rm Chr}^2=0.5$ (measured using the method described in 
Appendix A of 
Ref.~\cite{Lovelace2008}), we choose 
$M_{\rm ADM}\Omega_r=\mp 0.244146$ and 
$\xi^i = \partial_\phi$, where $\partial_\phi$ is the 
rotation vector on the apparent horizon corresponding to rotation about the 
$+z$ axis.

On the outer boundary $\mathcal{B}$, the spacetime metric is flat:
\begin{eqnarray}
\psi & = & 1 \mbox{ on }\mathcal{B},\\
\alpha\psi & = & 1 \mbox{ on }\mathcal{B}.
\end{eqnarray} Because the holes are initially at rest in the coordinates, 
they can be given orbital, radial, and translational motion by 
rotation, expansion, and translation of the shift on the outer boundary, 
i.e.
\begin{eqnarray}\label{eq:IDShiftOBBC}
\beta^i = \left(\mathbf \Omega_0 \times r\right)^i + \dot{a}_0 r^i + V_0^i, 
\mbox{ on }\mathcal{B}.
\end{eqnarray} We choose $\dot{a}_0=0$ and $\Omega$ = 0. To make 
the total momentum of the initial data vanish, we choose 
$V^i=-0.001444$.

Our initial data are 
constructed [Eq.~(\ref{eq:IDShiftBC})]
in a frame comoving
with the black holes. Thus, an asymptotic 
rotation, expansion, and translation in the comoving shift 
$\beta^i$ cause the holes to 
initially have radial, angular, or translational velocity in the
inertial frame. Note 
that the initial data are evolved in inertial, not comoving, 
coordinates, so that 
the shift during the evolution
is different from the comoving shift $\beta^i$ 
obtained from the XCTS equations: the former
asymptotically 
approaches zero, not a constant vector $V_0^i$.

\section{Numerical methods for evolutions}\label{sec:numer-meth-evol}
\subsection{Pseudospectral evolutions}\label{sec:pseud-evol}

We evolve the initial data summarized in Sec.~\ref{sec:IDSpEC} using 
the Caltech-Cornell pseudospectral code {\sc SpEC}.
This code and the methods it employs are described in detail
in Refs.~\cite{Scheel2006,Boyle2007,Scheel2008}.   Some of these methods
have been simplified for the head-on problem discussed here,
and others have been modified to account
for a nonzero center-of-mass velocity, so we will describe them here.

We evolve a first-order representation~\cite{Lindblom2006} of the
generalized harmonic
system~\cite{Friedrich1985,Garfinkle2002,Pretorius2005c}.  We handle
the singularities by excising the black hole interiors from the
computational domain. Our outer boundary
conditions~\cite{Lindblom2006,Rinne2006,Rinne2007} are designed to
prevent the influx of unphysical constraint
violations~\cite{Stewart1998,FriedrichNagy1999,Bardeen2002,Szilagyi2002,%
  Calabrese2003,Szilagyi2003,Kidder2005} and undesired incoming
gravitational radiation~\cite{Buchman2006,Buchman2007} while allowing
outgoing gravitational radiation to pass freely through the boundary.

We employ the dual-frame method described in Ref.~\cite{Scheel2006}:
we solve the equations in an ``inertial frame'' that is asymptotically
Minkowski, but our domain decomposition is fixed in a ``comoving
frame'' that is allowed to shrink, translate and distort relative to
the inertial frame.  The positions of the centers of the black holes
are fixed in the
comoving frame; we account for the motion of the holes by dynamically
adjusting the coordinate mapping between the two frames.  Note that
the comoving frame is referenced only internally in the code as a
means of treating moving holes with a fixed domain. Therefore all
coordinate quantities (e.g. black hole trajectories) mentioned in this
paper are inertial-frame values unless explicitly stated otherwise.

The mapping from comoving to inertial coordinates is changed
several times during the run. During the plunge phase, we denote
the mapping by ${\cal M}_{\rm p}(x^i,x'^i)$, where primed coordinates
denote the comoving frame and unprimed coordinates denote the inertial frame.
Explicitly, ${\cal M}_{\rm p}(x^i,x'^i)$ is the mapping
\begin{eqnarray}
\label{eq:CubicScaleMap}
x      &=& F(r',t) \sin\theta'\cos\phi', \\
\label{eq:CubicScaleMapa}
y      &=& F(r',t) \sin\theta'\sin\phi' + e^{-r'^2/r'^2_{\rm T}}Y(t), \\
\label{eq:CubicScaleMapb}
z      &=& F(r',t) \cos\theta'\cos\phi', 
\end{eqnarray}
where
\begin{eqnarray}
\label{eq:CubicScaleMapF}
  F(r',t) & := &
r'\left[a(t) + \left(1-a(t)\right) \frac{r'^2}{R_0'^2}\right].
\end{eqnarray}
Here $a(t)$ and $Y(t)$ are functions of time, $(r',\theta',\phi')$ are
spherical polar coordinates in the comoving frame centered at the origin,
and $R_0'$ and
$r'_{\rm T}$ are constants.  For the choice $R_0'=\infty$ and
$r'_{\rm T}=\infty$, the mapping is simply an 
overall contraction by $a(t)\le 1$ plus a translation $Y(t)$ 
in the $y$ direction.  Choosing $R_0'$ equal to the outer boundary 
radius~$R'_{\rm max}$
and choosing $r'_{\rm T} \sim R'_{\rm max}/6$ causes
the map to approach the identity near the outer boundary; this prevents
the outer boundary from falling close to the strong-field region during merger,
and makes it easier to keep the outer boundary motion smooth
through the merger/ringdown transition.
The functions $a(t)$ and $Y(t)$ are determined by dynamical control
systems as described in Ref.~\cite{Scheel2006}. These control systems
adjust $a(t)$ and $Y(t)$ so that the centers of the
apparent horizons remain stationary in the comoving frame.
For the evolutions presented here, we use 
$R_0'=532.2 M_{\rm ADM}=1.1 R'_{\rm max}$ 
and 
$r'_{\rm T}=31.21 M_{\rm ADM} = 4 d_o$, where $d_o$ is the initial separation 
of the holes.

The gauge freedom in the generalized harmonic system is fixed
via a freely specifiable gauge source function $H_a$ that satisfies the
constraint 
\begin{equation}
  \label{e:ghconstr}
  0 = \mathcal{C}_a := \Gamma_{ab}{}^b + H_a,
\end{equation}
where $\Gamma^{a}{}_{bc}$ are the spacetime Christoffel symbols.  To
choose this gauge source function, we define a new quantity
$\tilde{H}_a$ that transforms like a tensor and agrees with $H_a$ in
inertial coordinates (i.e. $\tilde{H}_a = H_a$).  Then we choose
$\tilde{H}_{a}$ so that the constraint~(\ref{e:ghconstr}) is satisfied
initially, and we demand that $\tilde{H}_{a'}$ is constant in the moving frame.

Shortly before merger (at time $t_1=31.1 M_{\rm ADM}$), we make two modifications
to our algorithm to reduce numerical errors and gauge dynamics during
merger.  First, we begin controlling the size of the individual apparent
horizons so that they remain constant in the comoving frame, and therefore
they remain close to their respective excision boundaries.  
This is accomplished by changing the
map between comoving and inertial coordinates as follows.  We define
the map ${\cal M}_{\rm AH_1}(\tilde{x}^i,x'^i)$ for black hole 1 as
\begin{eqnarray}
\tilde{x} &=& x'_{\rm AH_1} + \bar{r} \sin\theta'\cos\phi',\\
\tilde{y} &=& y'_{\rm AH_1} + \bar{r} \sin\theta'\sin\phi',\\
\tilde{z} &=& z'_{\rm AH_1} + \bar{r} \cos\theta',\\
\bar{r}   &:=& r' - e^{-(r'-r'_0)^3/\sigma_1^3} \lambda_1(t),
\end{eqnarray}
where $(r',\theta',\phi')$ are spherical polar coordinates
centered at the (fixed) comoving-coordinate location of black hole
1, which we denote as $(x'_{\rm AH_1},y'_{\rm AH_1},z'_{\rm AH_1})$.
The constant
$R'_{\rm AH_1}$ is the desired average radius (in comoving coordinates)
of black hole 1. Similarly, we
define the map ${\cal M}_{\rm AH_2}(\tilde{x}^i,x'^i)$ for black hole 2.
Then the full map from the comoving coordinates $x'^i$ to the inertial
coordinates $x^i$ is given by
\begin{equation}
  {\cal M}_{\rm m}(x^i,x'^i) := 
     {\cal M}_{\rm p}(x^i,\bar{x}^i)
     {\cal M}_{\rm AH_2}(\bar{x}^i,\tilde{x}^i)
     {\cal M}_{\rm AH_1}(\tilde{x}^i,x'^i).
\end{equation}
The constants $\sigma_1$, $\sigma_2$, and $r'_0$ are chosen to be 
$0.780 M_{\rm ADM}$,
$0.780 M_{\rm ADM}$, and $1.01 M_{\rm ADM}$, respectively.  
The functions $\lambda_1(t)$ and $\lambda_2(t)$
are determined by dynamical control systems that drive the comoving-coordinate 
radius of the apparent horizons towards their desired values 
$R'_{\rm AH_1}=R'_{\rm AH_2}=1.56M_{\rm ADM}$  
Note that in comoving coordinates, the shape of
the horizons is not necessarily spherical; only the average radius of the
horizons is controlled.

The second change we make at time $t_1=31.1 M_{\rm ADM}$ is to
smoothly roll gauge source function $H_a$ to zero by
adjusting $\tilde{H}_{a'}(t)$ according to
\begin{eqnarray}
\label{eq:gaugehfalloff}
\tilde{H}_{a'}(t) =\tilde{H}_{a'}(t_1) e^{-(t-t_1)^2/\tau^2},
\end{eqnarray} where $\tau=0.5853 M_{\rm ADM}.$ 
This choice makes it easier for us to 
continue the evolution after the common horizon has formed, and it also
reduces gauge dynamics that otherwise cause oscillations in the observed 
Landau-Lifshitz velocity $v^y_{\rm LL}$ during the ringdown.

When the two black holes are sufficiently close to one another, a new
apparent horizon suddenly appears, encompassing both black holes.
At time $t_m =34.73 M_{\rm ADM}$ 
(which is shortly after the common horizon forms),
we interpolate all
variables onto a new computational domain that contains only a single
excised region, and we choose a new comoving coordinate system so that
the merged (distorted, pulsating) apparent horizon remains spherical
in the new comoving frame.  This is accomplished in the same way as
described in Section II.D. of~\cite{Scheel2008}, except that here the
map from the new comoving coordinates to the inertial coordinates
contains an additional translation in the $y$ direction that handles
the nonzero velocity of the merged black hole.  In~\cite{Scheel2008} a
third change, namely a change of gauge, was necessary to continue the
simulation after merger. But in the simulations discussed here,
Eq.~(\ref{eq:gaugehfalloff}) has caused $H_a$ to fall to zero by the
time of merger, and we find it suffices to simply allow $H_a$ to
remain zero after merger.

For completeness, we now explicitly describe the map from the new comoving
coordinates $x''^i$ to the inertial coordinates $x^i$. This map is
given by
\begin{eqnarray}
\label{eq:PostMergerMap}
x         &=&  r \sin\theta''\cos\phi'',\\
y         &=&  r \sin\theta''\sin\phi'' + e^{-r''^2/r''^2_{\rm T}}Y(t),\\
z         &=&  r \cos\theta'',\\
r         &=&  \tilde{r}\biggr[1+\sin^2(\pi\tilde{r}/2 R''_{\rm max})  
                    \nonumber \\
                    &&\left. \times
                    \left(A(t)\frac{R'_{\rm max}}{R_{\rm max}''} 
                       + (1-A(t))\frac{R_{\rm max}'^3}{R_{\rm max}''R_0'^2} 
                       -1\right) \right], \\
\label{eq:PostMergerMapDistort}
\tilde{r} &=& r'' - q(r'')\sum_{\ell=0}^{\ell_{\rm max}} 
                       \sum_{m=-\ell}^{\ell} \lambda_{\ell m}(t)
                       Y_{\ell m}(\theta'',\phi''),
\end{eqnarray}
$(r'',\theta'',\phi'')$ are spherical
polar coordinates in the new comoving coordinate system, $R''_{\rm max}$ 
is the value of $r''$ at the outer boundary, and $r''_{\rm T}$ is a constant
chosen to be $31.21 M_{\rm ADM}$.  The function $q(r'')$ is given by
\begin{equation}
q(r'') = e^{-(r''-R''_{\rm AH})^3/\sigma_q^3},
\end{equation}
where $R''_{\rm AH}$ is the desired
radius of the common apparent horizon in comoving
coordinates.
The function $A(t)$ is 
\begin{equation}
A(t) = A_0 + (A_1+A_2(t-t_m))e^{-(t-t_m)/\tau_A},
\end{equation}
where the constants $A_0$, $A_1$, and $A_2$ are chosen so that
$A(t)$ matches smoothly onto $a(t)$ from Eq.~(\ref{eq:CubicScaleMapF}):
$A(t_m)=a(t_m)$, $\dot{A}(t_m)=\dot{a}(t_m)$, and $\ddot{A}(t_m)=\ddot{a}(t_m)$.
The constant $\tau_A$ is chosen to be on the order of $5M$.
The functions $Y(t)$ and $\lambda_{\ell m}(t)$ are determined by dynamical
control systems that keep the apparent horizon spherical and centered at the
origin in comoving coordinates;  see~\cite{Scheel2008} for details.

\subsection{Moving-puncture evolutions}
\label{app: movpunc}
In addition to the spectral evolutions, we have performed
a second set of simulations using the so-called {\em moving puncture}
technique \cite{Baker2006a, Campanelli2006a} using the {\sc Lean}
code \cite{Sperhake2006, Sperhake2007}. This code is based on the
{\sc Cactus} computational toolkit \cite{CACTUS}
and uses mesh refinement
provided by the {\sc Carpet} package \cite{SchnetterHawleyHawke2004,
Carpetweb}. Initial data are provided in the form of the
{\sc TwoPunctures} thorn by Ansorg's spectral solver
\cite{AnsorgBruegmann2004} and apparent horizons are calculated
with Thornburg's {\sc AHFinderDirect} \cite{Thornburg1996, Thornburg2004}.

The most important
ingredient in this method for the present discussion is the
choice of coordinate conditions. A detailed study of alternative
gauge conditions in the context of moving puncture type black-hole
evolutions is given in Ref.~\cite{vanMeter2006}. In particular, they
demonstrate how the common choice of a second order in time evolution
equation for the shift vector $\beta^i$ can be integrated in time
analytically and thus reduced to a first order equation. Various test
simulations performed with the {\sc Lean} code
confirm their Eq.~(26) as the most efficient method to evolve the
shift vector. In contrast to the shift, moving puncture codes show
little variation in the evolution of the lapse function. Here we follow
the most common choice so that our gauge conditions are given by
\begin{eqnarray}\label{eq:PunctureLapseGauge}
  \partial_t \alpha &=& \beta^i \partial_i \alpha - 2\alpha K, \\[10pt]
  \partial_t \beta^i &=& \beta^m \partial_m \beta^i
       + \frac{3}{4}\tilde{\Gamma}^i - \eta \beta^i.
       \label{eq: shift}
\end{eqnarray}
$\tilde{\Gamma}^i$ is the contracted Christoffel symbol of the
conformal 3-metric, $K$ the trace of the extrinsic curvature
[see for example Eq.~(1) of \cite{Sperhake2006}] and $\eta$
a free parameter set to $1$ unless specified otherwise.
For further details about the moving puncture
method and the specific implementation in the {\sc Lean code} code
we refer to Sec.~II of Ref.~\cite{Sperhake2006}. Except for the
use of sixth instead of fourth order spatial discretization~\cite{Husa2007},
we did not find it necessary to apply any modifications relative
to the simulations presented in that work.

The calculation of the 4-momentum in the {\sc Lean} code is
performed in accordance with the relations listed in
Sec.~\ref{sec:LLformal}. The only difference is that in a
BSSN code the four metric and its derivatives are not directly
available but need to be expressed in terms of the 3-metric
$\gamma_{ij}$, the extrinsic curvature $K_{ij}$ as well as the
gauge variables lapse $\alpha$ and shift $\beta^i$. The key quantity
for the calculation of the 4-momentum is the integrand in
Eq.~(\ref{eq:MomTot}). A straightforward calculation
gives it in terms of the canonical ADM variables
\begin{eqnarray}
  \partial_{\alpha} H^{0\alpha 0 j} &=& \frac{1}{\chi^3} \left[
       \frac{3}{\chi} \gamma^{jm} \partial_m \chi + \gamma^{km}
       \gamma^{jn} \partial_k \gamma_{mn} \right], \\
  \partial_{\alpha} H^{i\alpha 0 j} &=& \frac{1}{\chi^3} \left[
       2\alpha (K^{ij}-\gamma^{ij}K) + \gamma^{ij} \partial_m \beta^m
       - \gamma^{im}\gamma_m \beta^j \right] \nonumber \\
       && - \beta^i \partial_{\alpha} H^{0\alpha 0 j},
\end{eqnarray}
where $K := K^i{}_i$ and $\chi := \det \gamma^{-1/3}$ have been
used for convenience because they are fundamental variables in our
BSSN implementation.


\section{Treatment of the Event Horizon}
\label{sec:EH}
In this subsection, we summarize the numerical methods used to 
find the event horizon in our pseudospectral simulations.

The event horizon of the merging black holes is determined by
the global structure of the spacetime, and thus identifying
its location on any given
  time slice requires knowledge of the full evolution of the
spacetime.  In order to determine the event horizon surface for our
pseudospectral evolutions, we use the ``geodesic
method'' implemented by Cohen, Pfeiffer and Scheel
\cite{CohenPfeiffer2008}, which locates the event horizon by
evolving null geodesics backwards in time.  This
  algorithm makes use of the well-established property that outgoing
  null geodesics in close proximity to the event horizon diverge
  exponentially from it, when followed forwards in time.  Thus, these
  geodesics, when followed backwards in time, converge exponentially
  onto the event horizon, as first recognized by Libson et. 
  al.~\cite{Libson95a,Libson96}. 
One must evolve many
geodesics to get a full picture of the horizon. 
In this subsection, we explain how
these geodesics are chosen.

The event horizon finding process can be summarized in three steps:
$1)$ choosing a suitable locus of geodesics such that when evolved
backward in time, they map out the event horizon,  $2)$ evolving those
geodesics backwards in time from a late enough time such that the
spacetime is no longer dynamical (i.e. after the
 merged hole has  
rung down to its final state), and $3)$
determining which of those geodesics are on the event horizon at any
given time.  

One property of event horizons is the formation of
  ``cusps'' in the course of a black hole merger.  As the holes
  approach merger, generators enter the horizon through these
  cusps~\cite{Shapiro1995}.  Therefore, as we follow the generators
  backwards in time, they will \textit{leave} the horizon, and become
future generators of the event horizon. This implies that at early
  times, the event horizon will be a \textit{subset} of the surface
  defined by the geodesics.

We choose our locus of geodesics from the apparent horizon
surface at a time $t_{\rm end}=78.0 M_{\rm ADM}$. 
If the coordinate location of a point on the surface is $q^i$,
we can expand the
surface in terms of scalar spherical harmonics as
\begin{equation}
q^i(t,u,v) = \sum^L_{\ell=0} \sum_{m=-\ell}^{\ell} \tilde{A}^i_{\ell m}(t)Y_{\ell m}(u,v),
\end{equation}
where $Y_{\ell m}(u,v)$ are the standard spherical harmonics,
though $u$ and $v$ are \textit{not} the standard spherical angular
coordinates; they are merely a conveniently chosen parameterization of
  the surface.

Individual geodesics are placed on a rectangular grid in $(u,v)$ of
dimension $L+1$ by $2(L+1)$, with the $u$ values chosen such
that $cos(u)$ are the roots of the Legendre polynomials of order L+1,
and the $v$ values are equally spaced in the interval $[0,2\pi]$
\cite{CohenPfeiffer2008}.
  Therefore, there are $N = 2(L+1)^2$ geodesics on 
this surface; we call
the value of $L$ the \textit{resolution} of the event horizon finding
run. 
The position of the geodesic in space is given as a 3-vector $q^i$
  which is evolved, along with its derivatives.  The initial velocity
  of the geodesics is chosen to be the outgoing null normal to the
  surface of the apparent horizon at $t_{\rm end} = 78.0 M_{\rm ADM}$.

The integration of null geodesics backwards in time is straightforward given 
the metric data from a simulation.  Writing the position of the null geodesic 
as $q^\mu$ (where $q^0:=t$), one can reexpress the geodesic equation in terms of coordinate 
time as
\begin{equation}
\ddot{q}^i = \Gamma^0_{\alpha\beta} \dot{q}^\alpha\dot{q}^\beta\dot{q}^i 
-  \Gamma^i_{\alpha\beta} \dot{q^\alpha}\dot{q}^\beta,
\end{equation}
where $\Gamma^\mu_{\alpha\beta}$ are the spacetime Christoffel symbols.  
Reexpressing this as a first order system, we have
\begin{subequations}
\begin{equation}
\dot{q}^i=p^i
\end{equation}
\begin{equation}
\dot{p}^i= \Gamma^0_{\alpha\beta}p^\alpha p^\beta p^i 
- \Gamma^i_{\alpha\beta} p^\alpha p^\beta .
\end{equation}
\end{subequations}

Finally, we must determine at what times some of the geodesics
  pass through the cusps of the event horizon, and leave the horizon
  (as we follow them backwards in time).   Our
  most useful tool for this is the surface area element of the event
  horizon.  This is defined as the square root of the determinant of
  the induced metric on the horizon $\sqrt{h}$, where 
\begin{equation}
h= \frac{1}{\sin^2u}\text{det} 
\left( \begin{array}{cc}
  \gamma_{ij}\partial_uq^i\partial_uq^j &\gamma_{ij}\partial_uq^i\partial_vq^j \\
  \gamma_{ij}\partial_vq^i\partial_uq^j & \gamma_{ij}\partial_vq^i\partial_vq^j \end{array}
\right),
\end{equation}
$\gamma_{ij}$ is the 3-metric, and the area of the event horizon is given by
\begin{equation}
A(t) = \int dA = \int \sqrt{h(t,u,v)}\sin u \,du \, dv.
\end{equation}

To determine
when a null geodesic leaves the event horizon,
(going backward in time), we note that all geodesics
leave the event horizon surface at a cusp. In our pseudospectral 
simulations, we observe that all cusps on the event horizon
 surface are also caustics.  Thus, our cusps may be identified
 by considering what happens to the area element of the
surface at a caustic: it goes to zero (see \S 4.4 of
\cite{CohenPfeiffer2008} for a more thorough discussion).  Thus, by
tracking the local area element $\sqrt{h}$ of each geodesic, we are
able to tell that it leaves the horizon at the time its area element
approaches zero.  If the local area element does not approach zero at
any time, then it represents a null generator that originated on one
of the two initial holes. 
We are then able to define a masking
function for each geodesic; this function tells us if the geodesic is on the
horizon at a given timestep. 

At this point we have located the event horizon surface at all times,
and thus we may calculate the surface integral of any quantity we wish on
the horizon.  We note one subtlety: the formation of cusps on the event
horizon surface introduces a problem for taking spectral derivatives
on the surface.  Thus, for calculating derivatives of quantities on
the event horizon surface, we use a 6th order finite differencing
stencil.  Note that this is an improvement on the 2nd order stencil used in
\cite{CohenPfeiffer2008}.

\section{Superposed-Harmonic-Kerr (SHK) initial data}
\label{sec:AppSHK}

We also present a simulation, H1 in Table~\ref{Table:SpECID}, that is
similar to S1 except that the initial separation between the holes is
larger and the gauge is nearly harmonic.  The construction of
this Superposed-Harmonic-Kerr initial data for this
run follows that of the Superposed-Kerr-Schild (S1) initial data
described in Appendix~\ref{sec:AppID}.  The differences are as
follows.

  The first difference is our choice of coordinates. In
  Appendix~\ref{sec:AppID}, the quantities $\gamma^a_{ij}$,
  $K_a$, and $\alpha_a$ that appear in 
  Eqs.~(\ref{eq:CMetricSKS}), (\ref{eq:TrKSKS}), and~(\ref{eq:IDLapseBC})
  refer to the three-metric, the trace of the extrinsic curvature, and 
  the lapse function of the
  Kerr metric in Kerr-Schild coordinates.  Here we still use
  Eqs.~(\ref{eq:CMetricSKS}), (\ref{eq:TrKSKS}), and~(\ref{eq:IDLapseBC}),
  but $\gamma^a_{ij}$,
  $K_a$, and $\alpha_a$ now refer to the 
  three-metric, the trace of the extrinsic curvature,
  and the lapse function
  of the Kerr metric in fully harmonic coordinates,
  Eqs. (22)-(31), (41) and (43) of Ref.~\cite{cook_scheel97}.
  Furthermore, the
  computational
  domain is excised on surfaces of constant Boyer-Lindquist radius,
  $r_{\textrm{BL}}$, where
  \begin{equation}
    \frac{x^2 + y^2}{{(r_{\textrm{BL}} - \tilde{M_a})}^2 
      + \tilde{S_a}^2/\tilde{M_a}^2}
    + \frac{z^2}{{(r_{\textrm{BL}} - \tilde{M_a})}^2} = 1 .
  \end{equation}

The initial coordinate separation was chosen to be $d=29.73
M_{\textrm{ADM}}$ and the Gaussian weighting parameter that appears in
Eqs.~(\ref{eq:CMetricSKS}), (\ref{eq:TrKSKS}),
and~(\ref{eq:IDLapseBC}) is $w_a=d/9$. To obtain $S/M_{\rm
  Chr}^2=\{0,0,\pm 0.5\}$ we choose $\Omega_r=\mp 0.261332 /
M_{\textrm{ADM}}$ in Eq.~(\ref{eq:IDShiftBC}), and to make the total
momentum vanish we choose $V_0^y = -0.0000582185$ in
Eq.~(\ref{eq:IDShiftOBBC}).

Solving the XCTS equations results in initial data that is
approximately harmonic. Harmonic coordinates satisfy 
$\nabla^{c}\nabla_{c} x^a =0$,
or equivalently, $\Gamma_a := \Gamma_{a b}{}^b = 0$.  
We can
evaluate the degree to which the harmonic gauge condition is satisfied
in our initial data by examining the normalized magnitude of
$\Gamma_a$:
 \begin{equation}
   f := {\left(\frac
     {\displaystyle \sum_a {\lvert \Gamma_a \rvert}^2}
     {\displaystyle \frac{1}{4} \sum_a \sum_b 
       {\lvert \Gamma_{a b}{}^b \rvert}^2}
     \right)}^{1/2}
   \label{e:normH}.
 \end{equation}

The denominator consists of the sum of squares of terms that must
cancel to produce $\Gamma_a=0$, so that $f=1$ corresponds to complete
violation of the harmonic coordinate condition.  On the apparent
horizons $f < 0.049$, while in the asymptotically flat region far from
the holes $f < 0.0083$.  In the regions where the Gaussians in
Eqs.~(\ref{eq:CMetricSKS}), (\ref{eq:TrKSKS}) and (\ref{eq:IDLapseBC})
transition the XCTS free data from harmonic Kerr to conformally flat we 
cannot expect the data to be strongly harmonic, and we find that $f < 0.12$.

The techniques employed in the spectral evolution from this SHK
initial data follow those used for the SKS initial data as described
in Appendix~\ref{sec:pseud-evol}.  In particular, the generalized harmonic
gauge source function, $H_a$ (Eq.~\ref{eq:SpECGauge}), is constructed by
demanding that $\tilde{H}_{a'}$ remains frozen to its value in the
initial data.  The evolution proceeds in nearly harmonic gauge because
of the way the initial data is constructed.

Three of these H1 evolutions were performed at resolutions of approximately
$61^3$,  $67^3$ and $72^3$ grid points.  
The constraints were found to be convergent.
The data presented in this paper is taken from the highest
resolution run.

These simulations are specifically constructed to provide data for comparison
with PN approximations, so we are restricted to remain in our 
approximately harmonic
gauge. However, currently this gauge choice prevents us from continuing 
our H1 evolutions
beyond the plunge phase; we have not observed the 
formation of a
common horizon.

\bibliography{References/References}
\end{document}